\documentclass[%
 reprint,
superscriptaddress,
 amsmath,amssymb,
 aps,
]{revtex4-2}
\usepackage{comment}
\usepackage{graphicx}
\usepackage{dcolumn}
\usepackage{bm}
\usepackage{hyperref}

\usepackage{braket}

\DeclareMathOperator{\sinc}{sinc}

\begin{document}

\preprint{APS/123-QED}

\title{
Strategies for generating separable photon triplets in waveguides and ring resonators}

\author{Gisell Lorena Osorio}
\email{gisell-lorena.osorio-osorio@polymtl.ca}
\affiliation{%
D\'epartement de g\'enie physique, \'Ecole Polytechnique de Montr\'eal, Montr\'eal, QC, H3T 1J4, Canada
}%
\author{Milica Bani\'{c}}%
\affiliation{%
National Research Council of Canada, 100 Sussex Drive, Ottawa, Ontario K1A 0R6, Canada
}%
\author{Nicol\'as Quesada}
\thanks{Author to whom any correspondence should be addressed}
\email{nicolas.quesada@polymtl.ca}
\affiliation{%
D\'epartement de g\'enie physique, \'Ecole Polytechnique de Montr\'eal, Montr\'eal, QC, H3T 1J4, Canada
}%

\date{\today}

\begin{abstract}
Photon triplet sources exhibit non-Gaussian features, a key property for applications in quantum computing and quantum information. However, spectral correlations can limit the performance and detection efficiency of these systems. Motivated by this observation, we present a theoretical analysis of the spectral properties of photon triplets generated through spontaneous third-order parametric down-conversion in photonic devices, and discuss strategies to quantify and minimize such correlations. We propose two approaches: dispersion engineering in waveguides and pump engineering in resonators. We apply these strategies in two realistic source designs, namely a high-index-contrast optical fiber and a silicon nitride microring resonator. Finally, we discuss detection strategies for probing non-Gaussian features of the triplet state. We find that it is feasible to achieve few-mode generation of photon triplets using state-of-the-art experimental systems, a crucial step toward practical applications of photon triplet sources in quantum technologies.
\end{abstract}

\maketitle


\section{\label{sec:introduction}Introduction}

The generation of multipartite entangled states has been a long-term objective in quantum optics and is a fundamental component of quantum information protocols and applications. Multiple approaches have been explored for their generation, including time-domain multiplexing of photon pairs \cite{meyer2022scalable,yokoyama2013ultra}, frequency-bin encoding \cite{banic2024integrated}, path encoding \cite{bergamasco2017generation,bao2023very}, among others \cite{ju2019creating, jia2025continuous}. Within this broad landscape, nonlinear optical processes stand out as flexible and scalable approaches for producing such states. For example, spontaneous parametric down-conversion (SPDC) and four-wave mixing have been reliable methods to generate photon pairs. As a result of recent advancements in the fabrication of nonlinear photonic devices and ultra-sensitive detection techniques \cite{caspani2017integrated, katiyi2025quantum, kong2024large, stasi2024enhanced}, the implementation of higher-order parametric processes is becoming increasingly feasible. An important subject in this area is the generation of photon triplets, either through cascaded second-order nonlinear optical processes \cite{hubel2010direct,krapick2016chip} or through third-order parametric down-conversion (TOPDC) \cite{chang2020observation,douady2004experimental,bertrand2025experimental}.

Triplets produced by TOPDC stand out for their distinctive characteristics, such as tripartite spectral entanglement, and non-Gaussian features \cite{braunstein1987generalized, agusti2020tripartite} including Wigner function negativity without postselection \cite{Walschaers2021nonGaussian}, non-Gaussian quadrature statistics  \cite{bencheikh2022demonstrating, zhang2021non}, non-Gaussian entanglement \cite{bencheikh2022demonstrating,agusti2020tripartite}, and signatures in higher-order quadrature moments that serve as indicators of non-Gaussianity \cite{zhang2023genuine}. These properties enable
 applications including heralded emission of photon pairs \cite{barz2010heralded}, direct generation of Greenberger-Horne-Zeilinger (GHZ)  states \cite{greenberger1990bell}, and third-order squeezing \cite{banaszek1997quantum}. A variety of implementations have been proposed, including nonresonant systems, such as optical fibers \cite{corona2011experimental, cavanna2020progress}, bulk crystals \cite{borshchevskaya2015three}, and waveguides \cite{banic2022resonant,moebius2016efficient,bencheikh2022demonstrating}; and resonant systems, such as ring resonators \cite{banic2022resonant} and superconducting parametric cavities \cite{chang2020observation}. 

While these earlier works have focused on designing sources with reasonable generation rates, little consideration has been given to the spectral characteristics of the generated light \cite{Chekhova2005Spectral}. In this matter, the photon triplet wavefunction is of particular interest, as it defines the spectral correlation properties of these quantum states. In many scenarios, such correlations are undesirable because they lead to highly multimode fields \cite{banic2024quantum, Graffitti2018design}: This \textit{multimodeness} restricts high-visibility interference, complicating detection through coherent methods (e.g., homodyne and heterodyne detection), and preventing applications that rely on interference, such as those involving optical quantum networks \cite{uren2005pure}. Furthermore, it limits the non-Gaussian features of the triplets' Wigner representation in a particular supermode \cite{ra2020non, banic2024quantum}, making these sources unsuitable for applications that rely on this characteristic \cite{lloyd1999quantum, weedbrook2012gaussian, Walschaers2021nonGaussian}.

While the quantification and minimization of spectral correlations in SPDC-based photon pair sources is a well-studied problem \cite{uren2005pure, Graffitti2018design}, similar approaches have not yet been proposed for the single-mode operation of TOPDC sources. In this work, we present a theoretical analysis of the spectral properties of photon triplets generated through TOPDC, and we propose strategies to quantify and minimize their spectral correlations. The paper is structured as follows: In Section \ref{sec:purity}, we introduce a measure to quantify spectral correlations based on the spectral decomposition of the single-photon reduced density matrix. We propose experimental strategies to control the photon triplets' spectral correlations through dispersion engineering in waveguides or optical fibers (Section \ref{sec:dispersion_engineering}), and pump engineering in ring resonators (Section \ref{sec:ring-resonators}). In Section \ref{sec:sample_calc}, we present examples of experimental proposals for generating factorable photon triplets, using the above-mentioned strategies. We provide sample calculations for high-index-contrast optical fibers in Section \ref{sec:optical_fibers}, and for a silicon nitride microring resonator in Section \ref{sec:resonators_results}. Finally, in Section \ref{sec:detection}, we discuss methods to measure distinctive features of nearly uncorrelated triphoton states using triple coincidence detection and homodyne measurements. Our conclusions are outlined in Section \ref{sec:conclusions}. 

\section{Spectrally uncorrelated photon triplet states}

In TOPDC, an input pump field with central frequency $\overline{\omega}_P$ propagates through a third-order nonlinear medium, leading to the probabilistic downconversion of a pump photon into three daughter photons with central frequencies $\overline{\omega}_{1}$, $\overline{\omega}_{2}$, and $\overline{\omega}_{3}$. In this parametric interaction, the energy conservation condition is satisfied, such that
\begin{equation}\label{eq:omegas}
    \overline{\omega}_P-\overline{\omega}_{1}-\overline{\omega}_{2}-\overline{\omega}_{3}=0. 
\end{equation}
We focus on the degenerate frequency configuration, in which the triplets are generated with identical central frequencies  $\overline{\omega}_F=\overline{\omega}_P/3$ and in the same spatial mode (labeled as $F$).

In a ``low-gain" regime, the quantum state resulting from this interaction takes the following form (see Appendix \ref{sec:quantum_state_both}) \cite{corona2011third, dot2012quantum, okoth2019seeded,dominguez2020third,banic2024quantum}:
\begin{equation} \label{eq:psi2}
    \ket{\phi} \approx \ket{\text{vac}}+\varepsilon \ket{\text{III}},
\end{equation}
where $|\varepsilon|^2$ is the probability of generating a photon triplet per pump pulse, and 
\begin{align} \label{eq:ket_III}
    \ket{\text{III}}=\frac{1}{\sqrt{6}} \int d\omega_1 d\omega_2 d\omega_3 \Psi (\omega_1,\omega_2,\omega_3)\nonumber \\ a^\dagger_{F}(\omega_1)a^\dagger_{F}(\omega_2) a^\dagger_{F}(\omega_3)\ket{\text{vac}}
\end{align}
is a three-photon state~\cite{rohde2007spectral}. The operator $a^\dagger_{F}(\omega_i)$ is a creation operator satisfying the standard bosonic commutation relations, $\omega_i$ are frequency variables, $\ket{\text{vac}}$ denotes the vacuum state, and $\Psi (\omega_1,\omega_2,\omega_3)$  represents the \textit{triphoton joint spectral amplitude} (JSA).  The triphoton JSA is normalized according to
\begin{equation} \label{eq:JSAnormalization}
    \int d\omega_1d\omega_2 d\omega_3 |\Psi\left(\omega_1,\omega_2,\omega_3\right)|^2=1,
\end{equation}
and it is our function of interest because it captures the spectral correlation properties of the generated photon triplets.
Its form is determined by the pump spectral distribution and the nonlinear medium dispersion characteristics  (See Appendix \ref{sec:quantum_state_both}).

\subsection{ \label{sec:purity} Separability of the triphoton joint spectral amplitude}

The quantification of spectral correlations in SPDC-based photon pair sources is a well-studied problem \cite{uren2005pure,Graffitti2018design}, and the necessary and sufficient conditions for their minimization are well-established~\cite{quesada2018gaussian,houde2024ultrashort}. Typically, the separability in those sources is determined by performing a Schmidt~\cite{law2000continuous} (or Autonne-Takagi~\cite{houde2024matrix}) decomposition over the bipartite JSA. The generalization of this strategy to tripartite systems is not straightforward, since there are no general Schmidt decompositions for them \cite{PERES1995higher} (although significant progress continues to be made in understanding the entanglement in these systems~\cite{florido2024product}). Moreover, for the cases we consider here -- namely the generation of degenerate triplets -- the characterization of identical particle entanglement is more nuanced than that of entanglement of distinguishable particles~\cite{killoran2014extracting,sciara2017universality}. 

We base our spectral separability measure on properties of the single-particle density matrix of the triplet
\begin{align}\label{eq:rhoI}
\rho_I = \int d\omega_1 d\omega'_1 \rho(\omega_1,\omega'_1) a^\dagger_F(\omega_1)\ket{\text{vac}}\bra{\text{vac}}a_F(\omega'_1),
\end{align}
where the frequency-space density matrix is given by
\begin{align} \label{eq:reduced_rho}
\rho(\omega_1,\omega_1')=\int d\omega_2 d\omega_3 \Psi\left(\omega_1,\omega_2,\omega_3\right)\Psi^\ast(\omega_1',\omega_2,\omega_3).
\end{align}
and they are both normalized according to $\text{Tr}(\rho_I)= \int d \omega \rho(\omega,\omega) = 1$. With the density matrix we can introduce a separability parameter as
\begin{align}\label{eq:kappadef}
\kappa = \frac{1}{\text{Tr}(\rho_I^2)},
\end{align}
in analogy to the Schmidt number in the bipartite case.
Note that because $\rho(\omega_1,\omega'_1) = \rho^*(\omega'_1,\omega_1)$ and $\int d \omega_1 \rho(\omega_1,\omega_1)=1$, one can write an eigendecomposition 
\begin{align}\label{eq:dm}
\rho(\omega_1,\omega'_1) = \sum_n r_n f_n(\omega_1) f^*_n(\omega'_1),
\end{align}
where the $f_n(\omega_1)$ -- which we call ``pseudo-Schmidt functions''-- are an orthonormal basis, $\sum_n r_n=1$, and 
\begin{equation}\label{eq:kappa}
    \kappa=\frac{\left(\sum_n r_n\right)^2}{\sum_n r_n^2} = \frac{1}{\sum_n r_n^2}.
\end{equation}

One can also show that Eq.~\eqref{eq:kappadef} constitutes a bona-fide measure of spectral (non-)separability by extending the arguments made by Ma et al.~\cite{ma2011measure} in reference to a permutationally symmetric tripartite state.\footnote{Since the JSA is permutationally invariant, all the reduced density matrices are equal and so is the trace of their square. In Ma et al.~\cite{ma2011measure} they use as measure of entanglement the concurrence by choosing $C(x)  = \sqrt{2(1-x)}$ with $x$ being the trace of the square of the reduced density matrix (which is bounded between 0 and 1). Instead, we use a different, but equally valid, monotonic function, namely $g(x)=1/x$ in the domain $0\leq x\leq 1$, arriving at Eq.~\eqref{eq:kappa}.}

With the above points in mind, we conclude that minimizing spectral correlations is equivalent to minimizing $\kappa$. Indeed, since $\sum_n r_n=1$, it follows that
$\kappa\geq 1$, and $\kappa=1$ if and only if only one eigenvalue $r_n$ is nonzero; this corresponds to a separable JSA. We also emphasize once more that the measure in Eq.~\eqref{eq:kappadef} captures the amount of spectral correlation in the JSA, not the amount of entanglement between the three indistinguishable photons in the state in Eq.~\eqref{eq:ket_III}.

The first-order coherence function~\cite{fabre2020modes} of the triplet can be written as
\begin{align}
\braket{\phi|a_F ^\dagger(\omega_1) a_F(\omega_1')|\phi} =  3 \varepsilon^2 \rho(\omega_1,\omega_1'),
\end{align}
giving a second physical interpretation to the reduced density matrix. Moreover, we note that other low order moments of the ladder operators are zero, namely $\braket{\phi|a_F (\omega_1) |\phi}=0$ and $\braket{\phi|a_F (\omega_1) a_F(\omega_1')|\phi}=0$.
Given these observations about the pure state $\ket{\phi}$, we can use the elegant results from Mele \emph{et al.}~\cite{mele2025symplectic} to obtain a simple measure of non-Gaussianity as given by the \emph{symplectic rank}. This quantity is given by the number of symplectic eigenvalues of the covariance matrix of a pure state with values higher than those associated with the vacuum state. The $n^{\text{th}}$ symplectic eigenvalue associated with $\ket{\phi}$ exceeds that of the vacuum state precisely by $3 \varepsilon^2 r_n$, giving another interpretation to the reduced density matrix of the triplet state.

\subsection{\label{sec:dispersion_engineering} Dispersion engineering strategy for waveguides}

For a waveguide, the triphoton JSA is the product of two functions:
\begin{equation}\label{eq:twf}
 \Psi  (\omega_1,\omega_2,\omega_3)= \overline{\Psi} _{0}~ \alpha(\omega_1+\omega_2+\omega_3)~\varphi(\omega_1,\omega_2,\omega_3),
\end{equation}
where $\alpha(\omega_1+\omega_2+\omega_3)$ is known as the pump envelope function (PEF), and $\varphi(\omega_1,\omega_2,\omega_3)$ is the phase matching function (PMF). By $\overline{\Psi} _{0}$ we denote a normalization constant (see Appendix \ref{sec:quantum_state_wg}).

The PEF represents the spectral distribution of the pump laser. Its profile can vary depending on the experimental conditions. For example, it may take the form of hyperbolic secant pulses, which are characteristic of optical solitons and pulses emitted from certain mode-locked lasers \cite{agrawal2019nonlinear}. More generally, the PEF can be tailored through pulse-shaping techniques. For simplicity, this work considers pulses with a Gaussian shape
\begin{equation}\label{eq:alpha}
    \alpha(\omega_1+\omega_2+\omega_3)=\overline{\alpha}_P \exp(-\tfrac{1}{2\sigma^2}(\omega_1+\omega_2+\omega_3-\overline{\omega}_{P})^2),
\end{equation}
where $\overline{\alpha}_P = \sqrt{\frac{N_p}{\sigma \sqrt{\pi}}}$ is a normalization constant, such that $\int d\omega |\alpha(\omega)|^2=N_P$, where $N_P$ is the average number of pump photons, and $\sigma$ represents the pump's bandwidth.

The PMF depends implicitly on the waveguide's geometry and material dispersion properties. For a waveguide that is homogeneous along the longitudinal direction, the PMF is defined as follows:
\begin{equation}\label{eq:phi}
    \varphi(\omega_1,\omega_2,\omega_3)=\sinc\left(\frac{\ell}{2} \Delta k (\omega_1,\omega_2,\omega_3)\right),
\end{equation}
where $\ell$ is the length of the nonlinear medium, and 
\begin{eqnarray}\label{eq:phase matching2}
    \Delta k (\omega_1,\omega_2,\omega_3)=&&  k_P(\omega_1+\omega_2+\omega_3)\nonumber \\&&-k_F(\omega_1)-k_F(\omega_2)-k_F(\omega_3)
\end{eqnarray}
is the phase mismatch for TOPDC, where $k_P(\omega)$ and $k_F(\omega)$ are the wavenumbers of the pump and triplet fields, respectively, evaluated at a frequency $\omega$. The wavenumbers are defined by the dispersion relation  
\begin{equation}
    k_{\scriptscriptstyle P,F}(\omega_i)=\frac{\omega_i \ n_{\scriptscriptstyle P,F}(\omega_i)}{c},
\end{equation} 
where $n_{\scriptscriptstyle P,F}(\omega_i)$ is the effective refractive index of the pump ($P$) or triplet ($F$) mode at the frequency $\omega_i$, and $c$ is the velocity of light in vacuum.

To analyze the effect of the waveguide dispersion parameters on the phase-mismatch, we perform a Taylor expansion of the dispersion relations around the central frequencies $\overline{\omega}_P$ or $\overline{\omega}_F$, working up to the second order and assuming that the higher-order dispersion terms, as well as  self- and cross-phase modulation (SPM and XPM) effects, are negligible. We have
\begin{align}\label{eq:delk}
    \Delta k(\omega_1,\omega_2,\omega_3)=&\left(\overline{k}_P-3\overline{k}_F \right)
    \nonumber \\ &+\left(\frac{1}{v_P}-\frac{1}{v_F}\right)\left(\delta \omega_1+\delta \omega_2+\delta \omega_3\right) \nonumber \\
    & +\tfrac{1}{2}\beta_P \left(\delta \omega_1+\delta \omega_2+\delta \omega_3\right)^2 \nonumber \\ 
    & -\tfrac{1}{2}\beta_F\left(\delta \omega_1^2+\delta \omega_2^2+\delta \omega_3^2\right),
\end{align}
where $\delta \omega_i=\omega_i-\overline{\omega}_F$, and $\overline{k}_P$ ($\overline{k}_F$), $v_P$ ($v_F$), and $\beta_P$ ($\beta_F$) are the wavenumber, group velocity and group velocity dispersion parameters evaluated at the pump (triplet) central frequencies, respectively. 

\begin{figure}[htbp]
    \centering
    \includegraphics[width=0.7\linewidth]{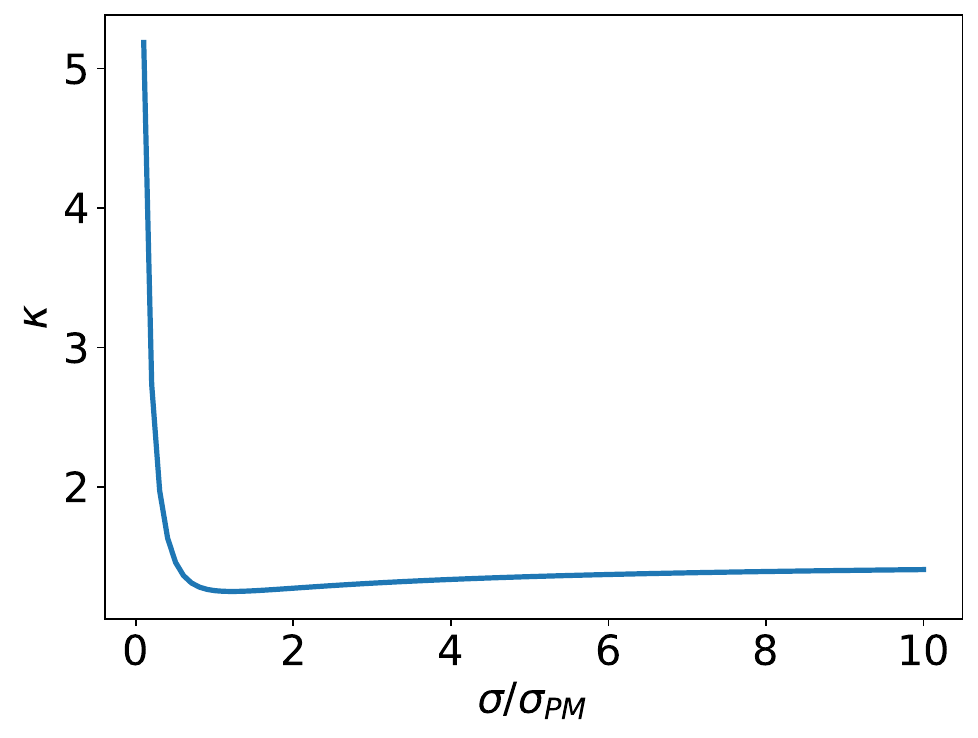}
    \caption{Separability parameter $\kappa$ for a waveguide in the ideal case where separability conditions are satisfied, shown as a function of the ratio between the pump pulse bandwidth $\sigma$ and the PMF bandwidth $\sigma_{PM}$. A minimum value of $\kappa\approx1.25$ is found near $\sigma\approx \sigma_{PM}$.}
    \label{fig:kappa_pump_bw}
\end{figure}

To achieve efficient TOPDC, one must satisfy phase matching $\left(\overline{k}_P-3\overline{k}_F=0 \right)$ and group-velocity matching $\left(v_P\approx v_F\right)$. Additionally, if $|\beta_F|\gg |\beta_P|$, the PMF can be approximated as 
\begin{equation}\label{eq:twf2}
    \varphi  (\omega_1,\omega_2,\omega_3) \approx  \sinc\left(\frac{\ell}{4} |\beta_F|\left(\delta \omega_1^2+\delta \omega_2^2+\delta \omega_3^2\right)\right),
   \end{equation}
which has isosurfaces corresponding to spherical layers in the three-dimensional frequency space $(\omega_1,\omega_2,\omega_3)$, resulting in a spherically symmetric spectral distribution (see Appendix \ref{sec:quadratic}).  It is then possible to achieve approximate separability of the JSA if, in addition to fulfilling the above conditions, the minimum pump bandwidth corresponds to the PMF bandwidth \cite{moebius2016efficient} 
\begin{equation}\label{eq:pmf_bw}
    \sigma_{PM} \propto \sqrt{4\pi/\ell|\beta_F|},
\end{equation}
as seen in Fig. \ref{fig:kappa_pump_bw}. This alignment ensures that the spherical symmetry of the PMF's main lobe is preserved. For this idealized scenario, we find a minimum separability parameter of $\kappa\approx 1.25$, establishing a theoretical lower bound for the separability parameter of dispersion-engineered TOPDC-based photon triplet sources. With this, we have established a set of conditions for approximate separability of the triphoton JSA for waveguides, and a lower limit for $\kappa$ for triplets generated in such systems. 

\begin{figure}
    \includegraphics[width=0.23\textwidth]{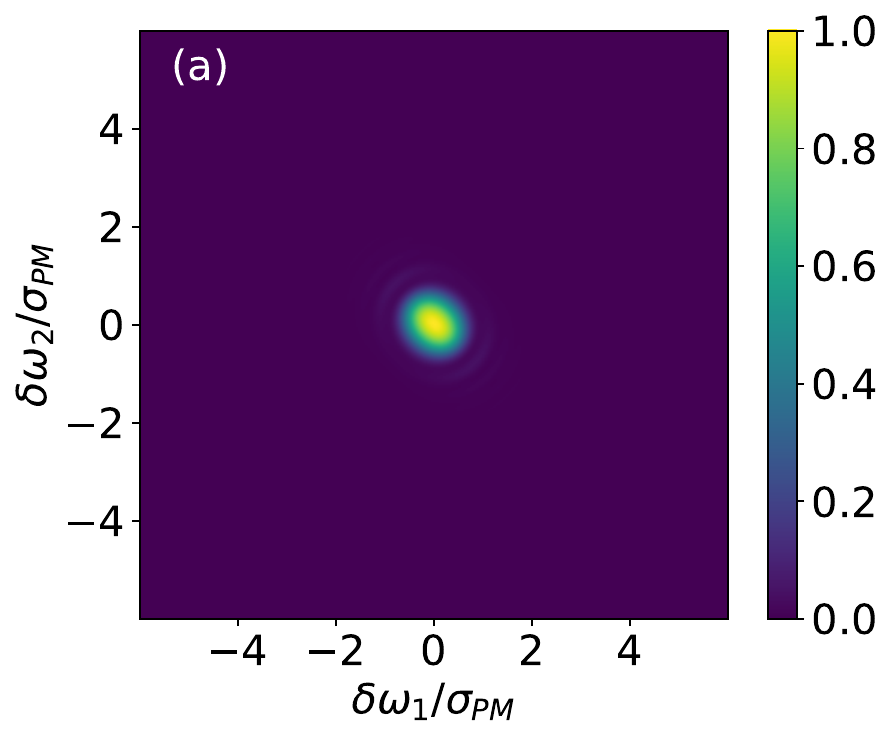}%
    \includegraphics[width=0.23\textwidth]{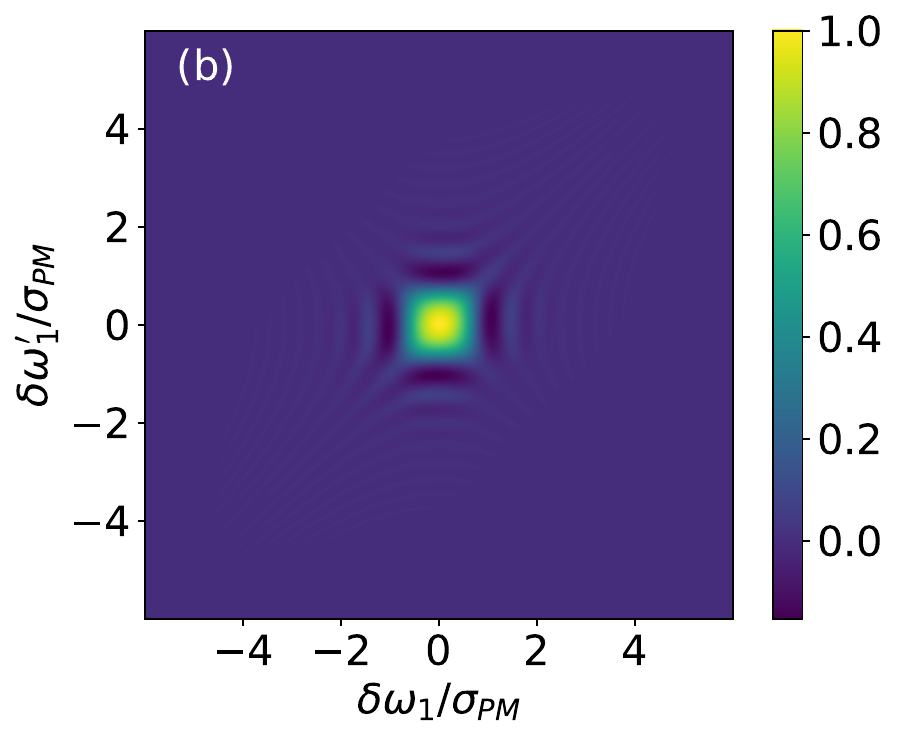}%
    \\
    \includegraphics[width=0.23\textwidth]{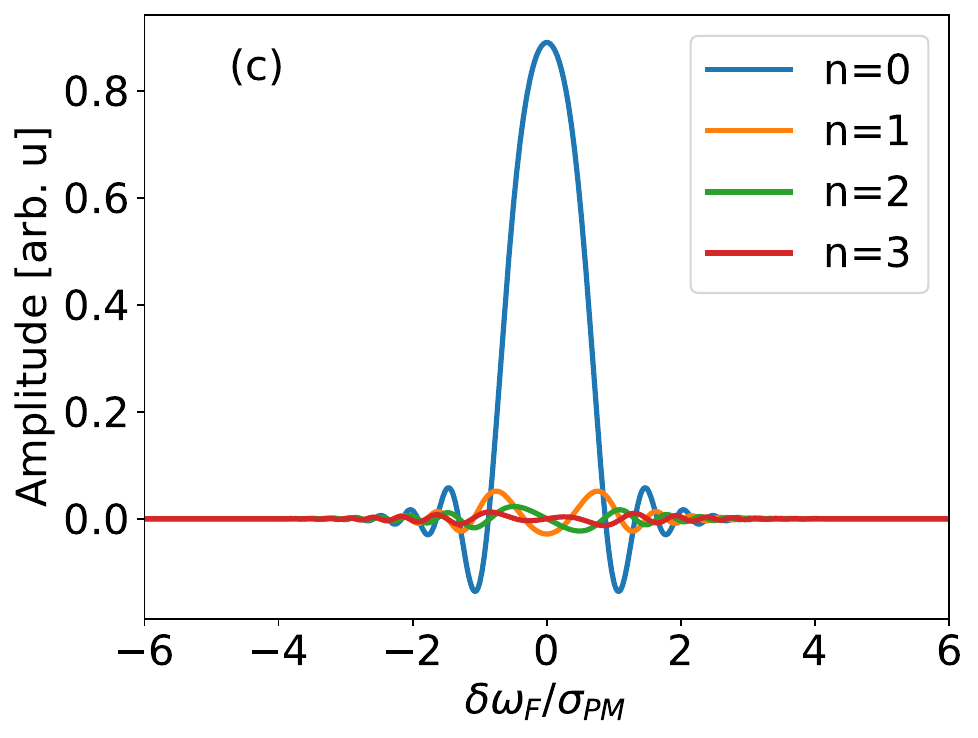}%
    \includegraphics[width=0.23\textwidth]{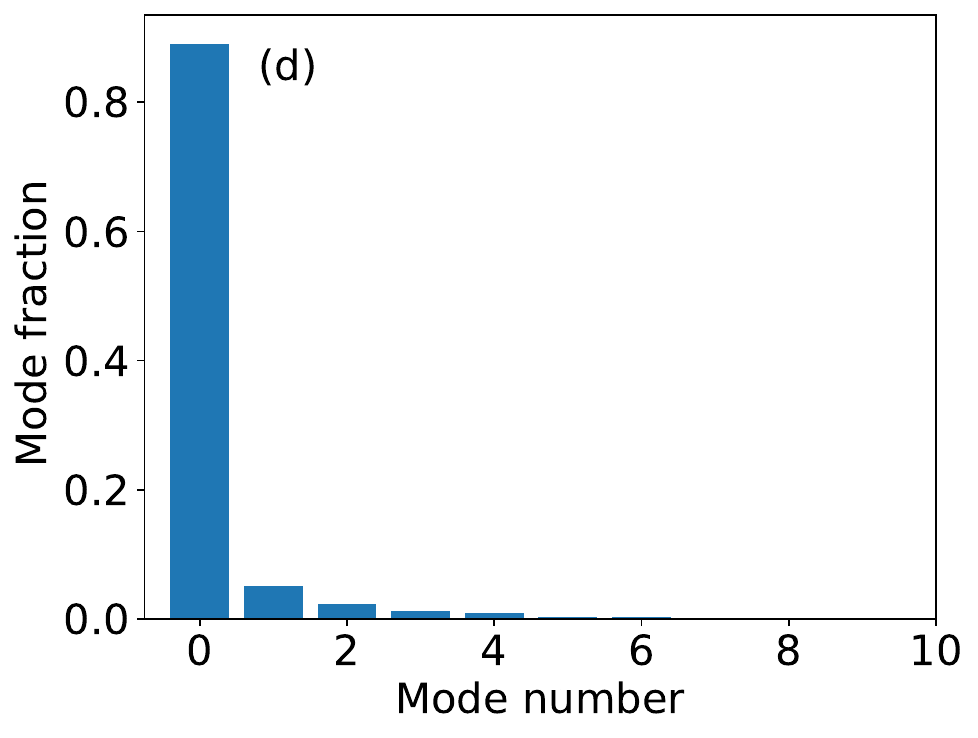}%
    \caption{\label{fig:jsirho_ideal} (a) Projection of the JSI $s(\omega_1,\omega_2)$ and (b) single-photon reduced density matrix $\rho(\omega_1, \omega_1')$ for the ideal approximated factorable case.(c) Dominant modes of $\rho(\omega_1, \omega_1')$, weighted by the mode fraction, and (d) mode distribution, highlighting the predominance of the first mode. }
\end{figure}

Fig. \ref{fig:jsirho_ideal} presents the spectral characteristics of a photon triplet source for the ideal case in which the separability conditions are fulfilled. To illustrate the symmetry argument discussed above, Fig. 2(a) shows a projection of the triphoton joint spectral intensity (JSI), $|\Psi(\omega_1,\omega_2,\omega_3)|^2$, onto the \{$\omega_1,\omega_2$\} plane, obtained by integrating over $\omega_3$: 
\begin{align}
    s(\omega_1,\omega_2)=\int d\omega_3 |\Psi(\omega_1,\omega_2,\omega_3)|^2. \label{eq:jsa_proj_def}
\end{align} 
The resulting distribution is approximately spherically symmetric. Identical representations are found for $s(\omega_1,\omega_3)$ and $s(\omega_2,\omega_3)$. The corresponding single-photon reduced density matrix $\rho_I$ is shown in panel (b). In (c), we present the dominant pseudo-Schmidt functions obtained from the singular value decomposition of $\rho_I$, weighted according to their mode fraction $r_n$ (see Eq. \eqref{eq:dm}), which is shown in (d), highlighting the predominance of the first mode.  

\subsection{\label{sec:ring-resonators} Pump engineering in ring resonators}

In this section, we consider TOPDC in a high-finesse microring resonator. Such a device can be appropriately treated using a Hamiltonian formulation, using a simple model for the ring-channel coupling  \cite{beyondphotonpairs, vernon_PhysRevA.91.053802, banic_PhysRevA.106.043707}. Using this treatment, the JSA for TOPDC is found to be  (see Appendix \ref{sec:quantum_state_ring})
\begin{align}
    \label{eq:JSA_ring} \Psi_{R}(\omega_1,\omega_2,\omega_3) = &\overline{\Psi}_0 F^*_{F+}(\omega_1) F^*_{F+}(\omega_2) F^*_{F+}(\omega_3)\\ &\times F_{P-}(\omega_1 + \omega_2 + \omega_3) \alpha (\omega_1+\omega_2+\omega_3), \nonumber
\end{align}
where $\alpha(\omega_1+\omega_2+\omega_3)$ is the pump envelope function as described above, $\overline{\Psi}_0$ is again a normalization constant, and 
\begin{align}
    F_{J\pm}(\omega) = \frac{1}{\sqrt{\mathcal{L}}} \left( \frac{\gamma^*_J}{(\omega_J - \omega) \pm i\overline{\Gamma}_J}\right) \label{eq:F_def}
\end{align}
is a field enhancement factor associated with the resonance labeled $J$. By $\mathcal{L}$ we denote the ring circumference, $\omega_J$ refers to the resonance frequency, $\overline{\Gamma}_J$ denotes the resonance linewidth, and $\gamma_J$ describes the coupling between the ring and the bus waveguide \cite{banic2022resonant, vernon_PhysRevA.91.053802}. The loaded quality factor of the resonator is defined in terms of the linewidth $\overline{\Gamma}_J$ as
\begin{align}
    Q_J = \frac{\omega_J}{2\overline{\Gamma}_J}.
\end{align}

A factorable JSA can be obtained from a ring resonator using the following approach, which has been applied for the generation of spectrally uncorrelated photon pairs \cite{Vernon:17}. Consider a scenario in which the resonances at $\omega_P$ and $\omega_F$ are mismatched, such that $\overline{\Gamma}_P > \overline{\Gamma}_{F}$, with the ring resonator pumped by a pulse shorter than the dwelling time $\tau_F \sim 1/\overline{\Gamma}_F$. In this limit, the range of frequencies contributing to the JSA is constrained (to $\sim \overline{\Gamma}_F$), while the functions $\alpha(\omega_1 + \omega_2 + \omega_3)$ and $F_{P-}(\omega_1 + \omega_2 + \omega_3)$ are broad in comparison, such that they vary relatively little over relevant values of $\omega_1 + \omega_2 + \omega_3$. In this case we can write $\alpha(\omega_1+\omega_2+\omega_3) \approx \overline{\alpha}_P$ (recall Eq. \ref{eq:alpha}), and similarly for the pump field enhancement factor; that is, we approximate the JSA as
\begin{align}
    \label{eq:phi_R_sep} \Psi_{R}(\omega_1,\omega_2,\omega_3) &\approx \overline{\Psi}_0 F^*_{F+}(\omega_1) F^*_{F+}(\omega_2) F^*_{F+}(\omega_3)\\ &\times F_{P-}(\omega_P) \overline{\alpha}_P , \nonumber 
\end{align}
which has the separable form $\Psi_{R}(\omega_1,\omega_2,\omega_3) \approx \psi(\omega_1)\psi(\omega_2)\psi(\omega_3)$ with 
\begin{align}
    \psi(\omega) \propto F^*_{F+}(\omega). \label{eq:lorentz_mode}
\end{align}
In this separable limit, the joint temporal amplitude (JTA) -- which provides insight into the dynamics of the triplet generation -- has a simple form. We have $\overline{\Psi}_R(t_1,t_2,t_3) \propto f(t_1)f(t_2)f(t_3)$, with 
\begin{align}
    f(t) &= i \gamma_F{\sqrt\frac{\mathcal{L}}{2\pi}} e^{-\overline{\Gamma}_Ft} e^{-i \omega_F t} \theta(t)
\end{align}
being the Fourier transform of $F^*_{F+}(\omega)$, where $\theta(t)$ denotes a Heaviside step function. 
A separable JSA (and JTA) can also be achieved by applying more elaborate pulse shaping schemes \cite{Christensen:18, Burridge:20},  but as above, we limit our discussion to Gaussian pump pulses.

\section{\label{sec:sample_calc} Sample calculations}

\subsection{\label{sec:optical_fibers} Generation of few-mode photon triplets in thin high-index-contrast optical fibers}

Optical fibers have been proposed as potential platforms for TOPDC. Earlier studies laid important groundwork, particularly in identifying viable designs in terms of phase-matching and generation rates \cite{corona2011experimental,corona2011third,cavanna2020progress,hammer2018dispersion}. Building on this foundation, we propose an experimental scheme for generating nearly uncorrelated photon triplets in thin high-index-contrast optical fibers. Unlike earlier approaches, our work focuses on the multimode structure of the triplet state, and addresses the separability of the triplet JSA when designing the optical fibers. 

Our scheme is inspired by the work of \citeauthor{tishchenko2024intermodal} \cite{tishchenko2024intermodal}, on fibers optimized for intermodal phase matching and group velocity matching for third harmonic generation (THG). We extend the approach of Ref. \cite{tishchenko2024intermodal} to target the dispersion conditions required for factorable TOPDC states, as outlined in Section \ref{sec:dispersion_engineering}. THG, the inverse process of TOPDC, satisfies the same phase matching condition and, notably, has been successfully demonstrated in optical fiber systems \cite{ha2021efficient,coillet2012third,bencheikh2012phase, akimov2003generation, nicacio1993third}. 

\begin{figure}[htp]
    \includegraphics[width=0.23\textwidth]{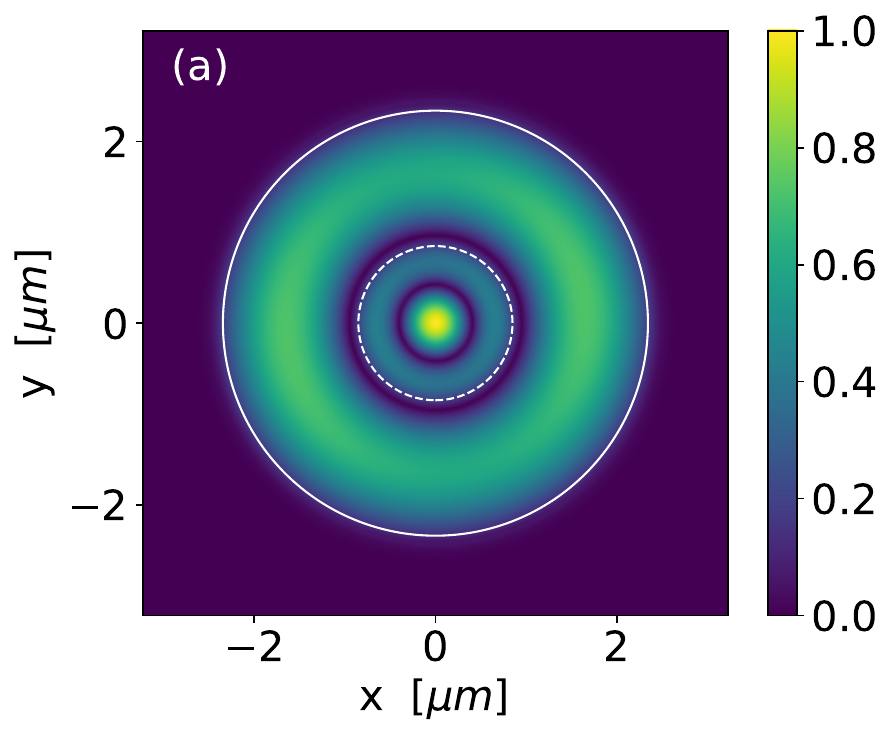}%
    \includegraphics[width=0.23\textwidth]{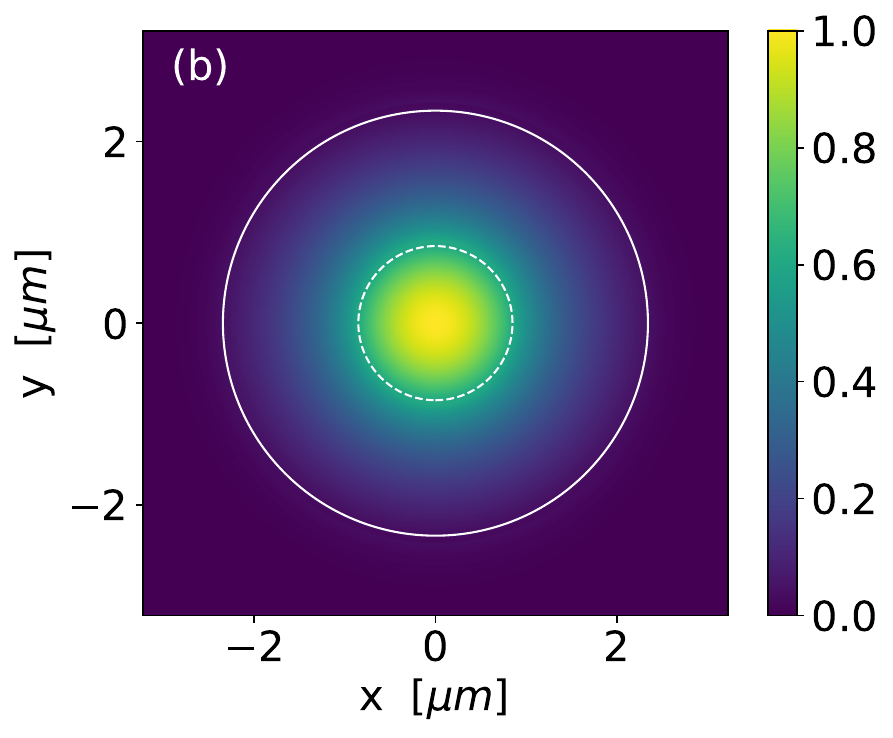}\\
    \vspace{1em}
    \includegraphics[width=0.4\textwidth]{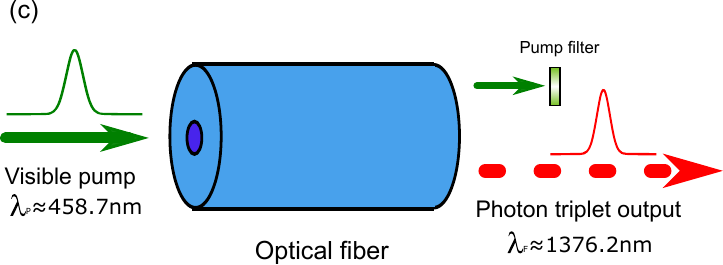}
    \caption{\label{fig:modes} Plot of the electric field intensities of the (a) pump mode (LP03) and (b) triplet mode (LP01) supported in the proposed optical fiber. Dotted circle: Boundary of the core ($r_{\text{core}}=0.849~\mu$m, material: GeO$_2$ doped silica). Solid circle: Boundary of the cladding ($r_{\text{clad}}=2.341~\mu$m, material: fused silica). (c) Schematic representation of photon triplet generation in an optical fiber.}
\end{figure}

In our scheme, the interaction occurs between the LP03 pump mode and the LP01 mode for triplets (shown in Fig. \ref{fig:modes} (a) and (b)), in optical fibers with a GeO$_2$-rich core and a pure silica cladding (see Fig. \ref{fig:modes} (c)). 
We perform simulations using COMSOL Multiphysics software (version 6.3) to obtain the transverse mode profiles and the dispersion curves of the modes of interest, varying the core and cladding radii of the fibers. For these simulations, we model the material properties of the cladding and core using the Sellmeier equations for silica \cite{malitson1965interspecimen} and GeO$_2$ doped silica glasses at various molar fractions of GeO$_2$ \cite{fleming1984dispersion}. Besides considering the design principles outlined in Section \ref{sec:dispersion_engineering} in our optimization, we aim to maintain a realistic design in which phase-matching occurs at optical frequencies, while ensuring that the fabrication dimensions remain practical.

\begin{figure}[htp]
    \includegraphics[width=0.7\linewidth]{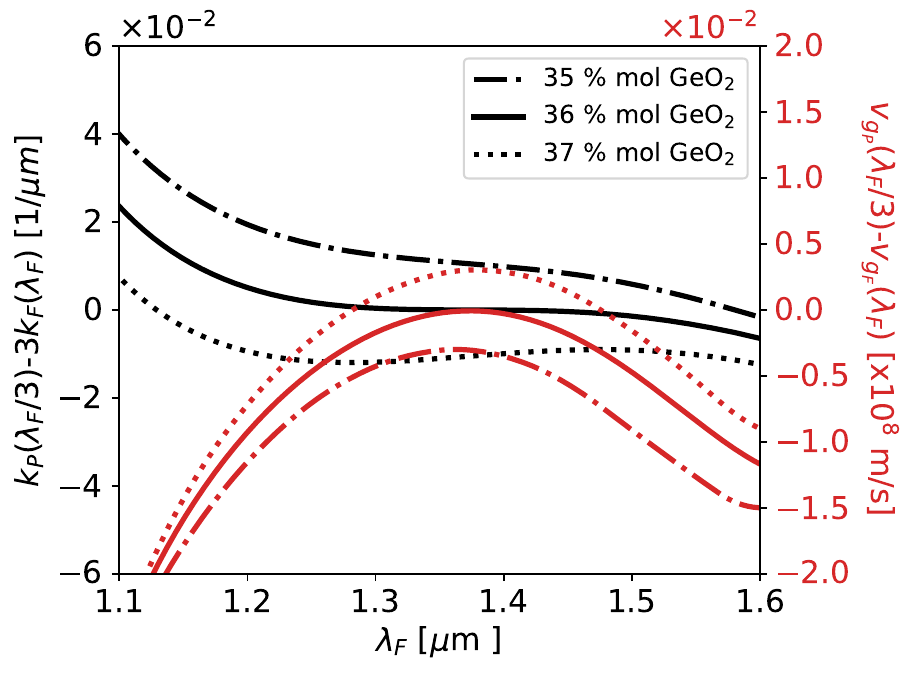}   \caption{\label{fig:pm_gvm} Phase-mismatch (black) and group-velocity mismatch (red) plotted as a function of the photon triplet wavelength ($\lambda_F$) in the degenerate frequency configuration for TOPDC in the optimized optical fiber taper ($r_{core}$=0.849 $\mu$m, $r_{clad}$=2.341 $\mu$m) at different GeO$_2$ doping levels. Simultaneous phase matching and group velocity matching are achieved at $\lambda_F \approx$  1.376 $\mu$m (equivalent to 13.69 $\times10^{14}$ rad/s) considering a 36\% mol GeO$_2$ doping level.}
\end{figure}

We find an optimal fiber design with a 36$\%$ mol GeO$_2$ doped silica core of 0.849 $\mu$m radius and a pure silica cladding of 2.341 $\mu$m radius. In this configuration, simultaneous intermodal phase matching and group-velocity matching are achieved for a pump tuned at $\lambda_P\approx$ 458.7 nm, leading to photon triplets generated at $\lambda_F\approx$ 1376.2 nm. For the pump and triplet modes we have $|\beta_P|$ = 6.4 fs$^2$/mm and $|\beta_F|$ = 21.9 fs$^2$/mm, respectively. In Fig. \ref{fig:pm_gvm}, we present the phase-matching (PM) and group-velocity-matching (GVM) curves in the degenerate frequency configuration for TOPDC as a function of the photon triplet wavelength and at various GeO$_2$ doping levels, considering the optimal fiber geometry mentioned previously. We can see in Fig. \ref{fig:pm_gvm} that varying the GeO$_2$ doping level alters the material dispersion, shifting both the PM and GVM curves. If the fiber geometry remains fixed, these shifts prevent simultaneous PM and GVM, necessitating re-optimization of the fiber design. Notably, we find that lower doping levels tend to result in slower group velocities for higher modes, making simultaneous PM and GVM more difficult to achieve. The optimal fibers exhibit phase-matching characteristics that are favorable for TOPDC, as the phase mismatch remains low over a broad spectral range. This is illustrated in Fig. \ref{fig:pm_gvm}, where the mismatch for the fiber with 36\% mol GeO$_2$ doping level remains on the order of $|\overline{k}_P-3\overline{k}_F| \approx 10^{-4}$ $\mu$m$^{-1}$ across the range $\lambda_F$ = 1.32-1.46 $\mu$m. 

Another important factor to consider is the loss of these materials; losses in high-GeO$_2$-doped fibers have been reported to be on the order of 2 dB/m at 450 nm and 0.01 dB/m at 1400 nm in 35$\%$ mol GeO$_2$ doped optical fibers \cite{dianov1997origin,tsvetkov2016phase}. While these losses are relatively high, this should not pose a major challenge for our platform: Due to the thin fiber dimensions, lengths are expected to be at most 30 cm to ensure structural stability. Given the short fiber lengths considered, we assume these losses are negligible, both for the fundamental triplet mode and higher-order pump mode. The thin optical fibers considered in this work can be fabricated through an adiabatic tapering process, with feasible geometries well established in the literature \cite{birks2002shape, azari2016theoretical,hou2024compact}.

\begin{figure}[htp]
    \includegraphics[height=3.6cm]{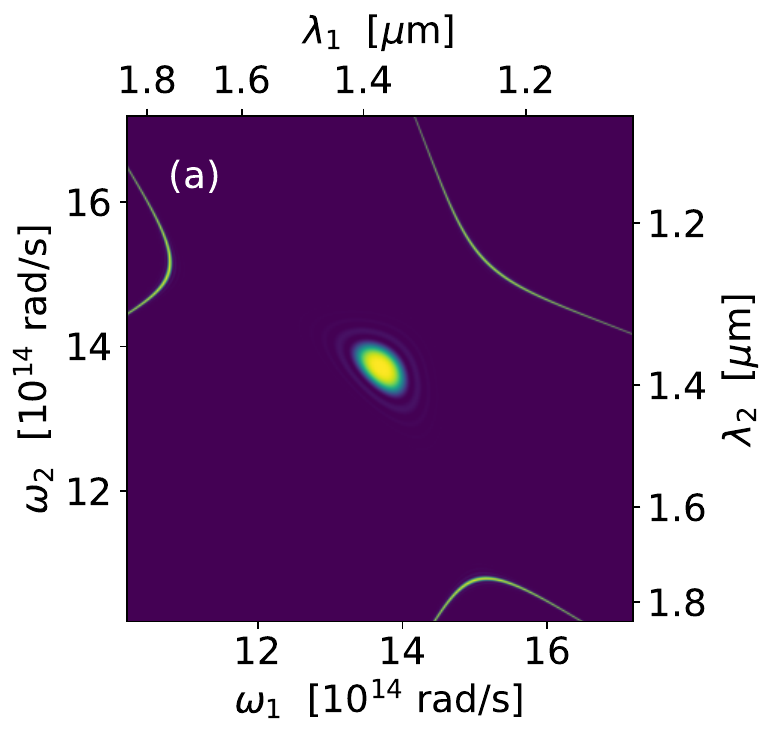}%
    \quad
    \includegraphics[height=3.6cm]{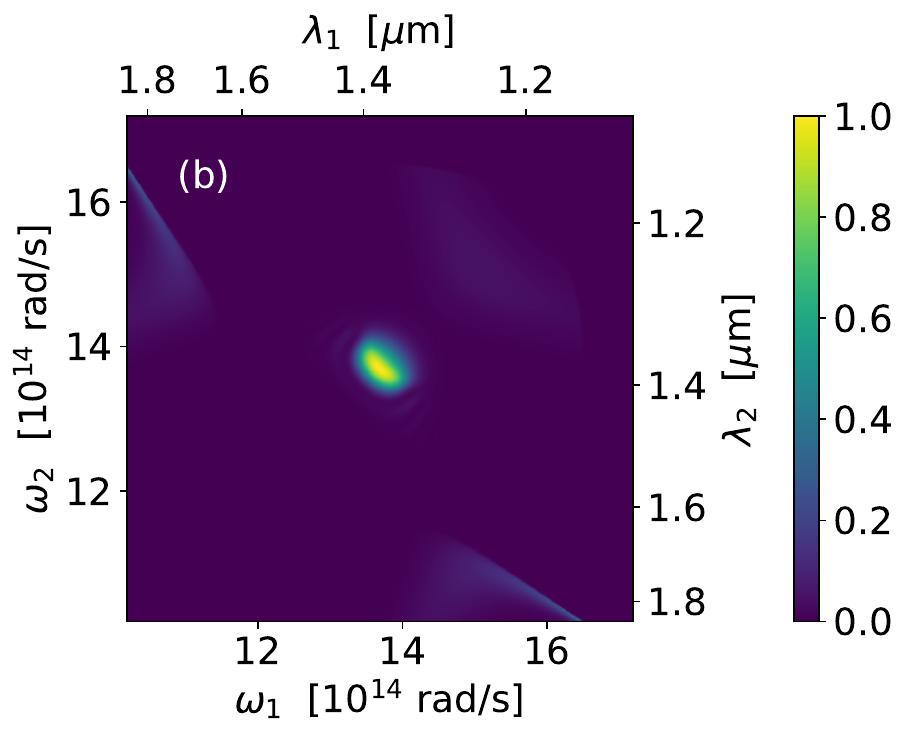}%
    \caption{\label{fig:unfiltered_jsa} Projections of the unfiltered JSI $s(\omega_1,\omega_2)$ for (a) continuous-wave (CW) and (b) pulsed pump configurations for the 36\% mol GeO$_2$-doped silica fibers described in Fig. \ref{fig:modes}. Corresponding wavelength axes ($\lambda_1,\lambda_2$) shown on the top and right. The central lobe corresponds to frequencies satisfying phase matching for degenerate triplet generation, while the surrounding side lobes indicate spectral regions supporting nondegenerate triplet generation. }
\end{figure}

We use the mode dispersion data to compute the triphoton JSA and the single-photon reduced density matrix, as outlined in Section \ref{sec:purity}.  Using Eq. \eqref{eq:pmf_bw} and assuming a fiber length of 20 cm, we estimate the PMF bandwidth as $\sigma_{PM}$ = 0.53 $\times10^{14}$ rad/s, corresponding to a pump pulse duration (FWHM) of 31 fs. For this particular fiber, the projection of the JSI $s(\omega_1,\omega_2)$ (recall Eq. \eqref{eq:jsa_proj_def}) exhibits a main lobe centered at $\omega_F$=13.69 $\times10^{14}$ rad/s (equivalent to 1.376 $\mu$m), corresponding to the phase-matching frequency for degenerate photon triplet generation. Additionally, {boomerang}-shaped side lobes are present, which indicate spectral regions where nondegenerate triplet generation is phase-matched and thus supported (see Fig. \ref{fig:unfiltered_jsa}). Since our focus is on the degenerate case, we apply spectral filtering to exclude frequencies below 11.5 $\times 10^{14}$ rad/s, and above 15 $\times 10^{14}$ rad/s.

\begin{figure}[ht]
    \includegraphics[width=0.23\textwidth]{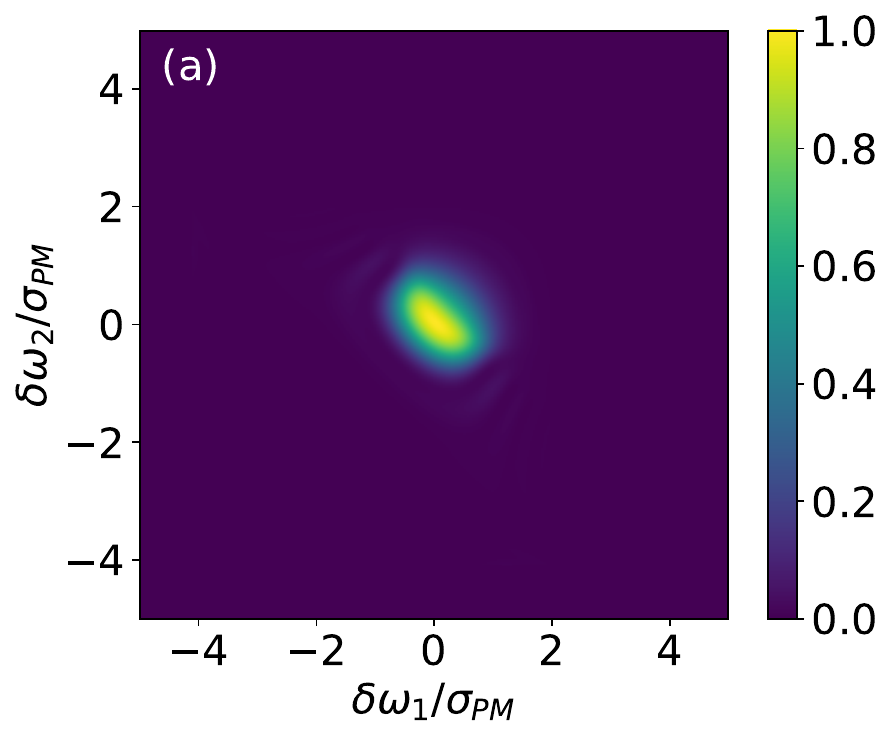}%
    \includegraphics[width=0.23\textwidth]{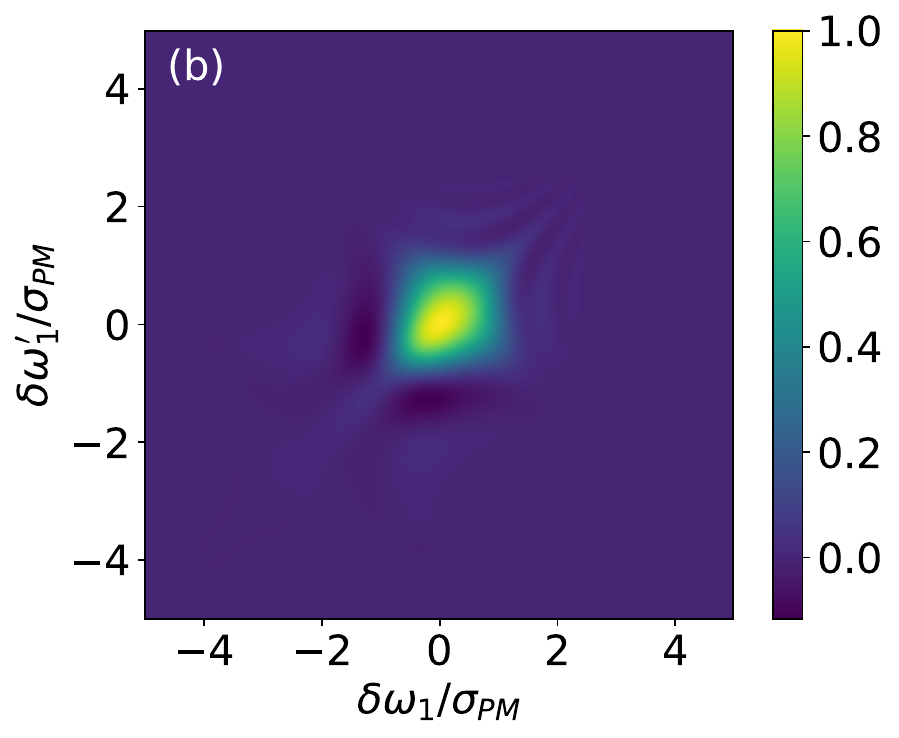}%
    \\
    \includegraphics[width=0.23\textwidth]{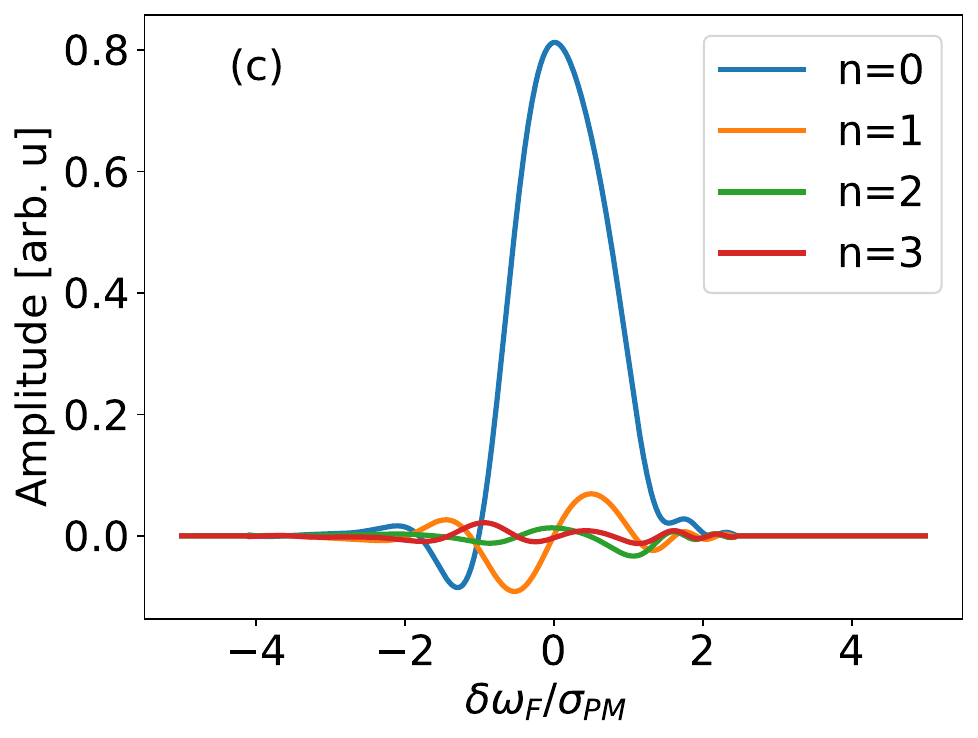}%
    \includegraphics[width=0.23\textwidth]{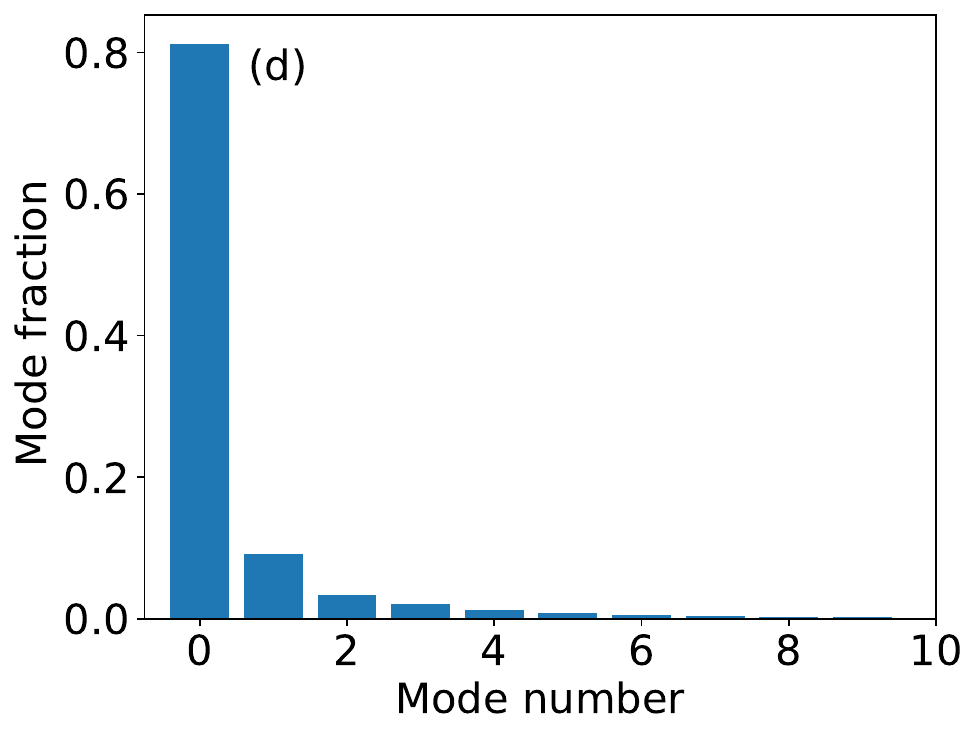}%
    \caption{\label{fig:jsirho} (a) Projection of the JSI $s(\omega_1,\omega_2)$, and (b) single-photon reduced density matrix $\rho(\omega_1, \omega_1')$, calculated using the dispersion profile of the optimal 36\% mol GeO$_2$-doped silica fibers described in Fig. \ref{fig:modes}. (d) Dominant modes of $\rho(\omega_1, \omega_1')$, weighted by the mode fraction, and (e) mode distribution. The results confirm the few-mode operation of the proposed triplet source, as evidenced by the presence of a small number of dominant modes.}
\end{figure}

 In Fig. \ref{fig:jsirho}(a), we show the projection of the filtered JSI $s(\omega_1,\omega_2)$, and the corresponding single-photon reduced density matrix is depicted in Fig. \ref{fig:jsirho}(b). In panel (c), we show the dominant modes in the singular value decomposition of the single-photon reduced density matrix $\rho_I$, weighted according to their mode fraction $r_n$, as introduced in Eq.\eqref{eq:dm}. In panel (d), we present the mode distribution of $\rho_I$, confirming the few-mode operation of the triplet source. For this system, we obtain a separability parameter of $\kappa\approx 1.5$, indicating that the system operates in the few-mode regime. 

\begin{figure}
    \includegraphics[width=0.7\linewidth]{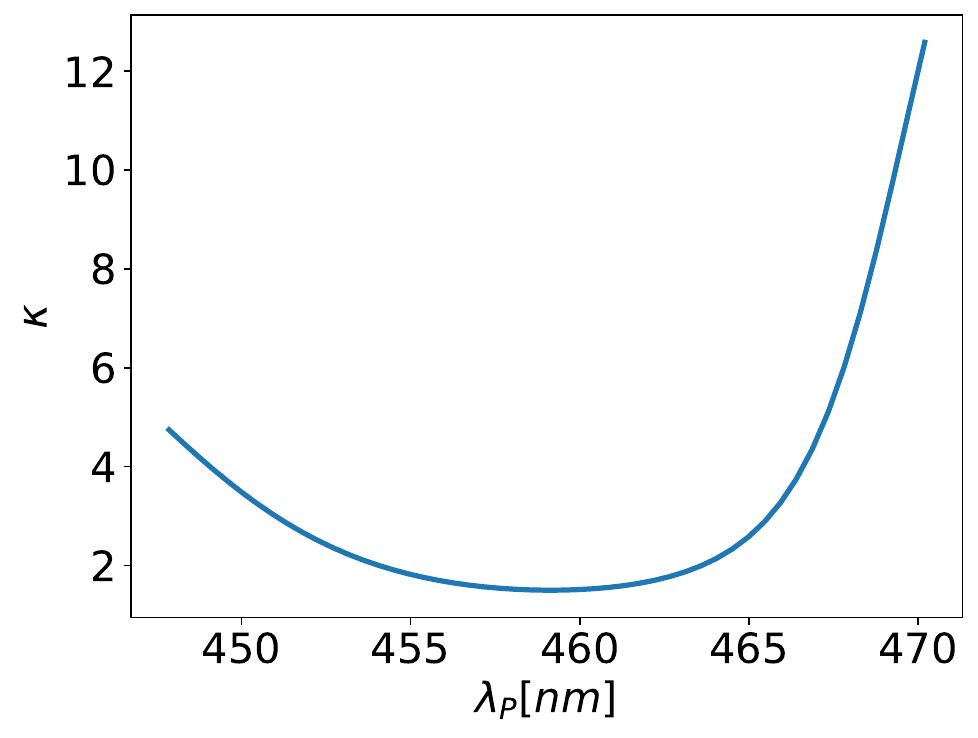}%
    \caption{\label{fig:pure_freq_sweep} Separability parameter $\kappa$ for the optimized fiber, plotted as a function of the pump central wavelength ($\lambda_P$) in the degenerate frequency configuration for TOPDC. The few-mode operation is maintained despite pump detunings up to 10 nm from the perfect phase-matching wavelength $\lambda_P\approx$ 458.7 nm.}
\end{figure}

 We now highlight a distinctive feature of these fibers: Triplet generation remains stable under variations in the pump frequency. Our design maintains single-digit values for $\kappa$, even with detuning the pump up to 10 nm from the perfect phase matching wavelength $\lambda_P\approx$ 458.7 nm (see Fig. \ref{fig:pure_freq_sweep}). This robustness is a consequence of the broad spectral range over which the phase mismatch remains low (as shown in Fig. \ref{fig:pm_gvm}).

Following the procedure described in Appendix \ref{sec:quantum_state_wg}, and assuming a pump average power of 50 mW, the predicted generation rate for this source is approximately 0.01 triplets/s. The triplet rate could be increased by extending the fiber length and raising the pump power, provided that SPM and XPM are effectively mitigated.  For instance, with a 30 cm fiber, we estimate a PMF bandwidth of 0.43 $\times10^{14}$ rad/s, corresponding to a 38 fs pump pulse. With this length and with an average pump power of 400 mW, the generation rate increases by an order of magnitude to 0.1 triplets/s, with a purity of about $\kappa \approx 1.5$; however, under these conditions, the pump may still be affected by SPM. One possible method to mitigate SPM consists of pre‑chirping the pump laser \cite{agrawal2019nonlinear,suda2012effects}. Further details are provided in Appendix \ref{sec:wg_SPM_XPM}. Although these rates are low and present a detection challenge, they approach the sensitivity limit of state-of-the-art single-photon detectors \cite{ID281}. The optimal triplet generation rate ultimately depends on the intended application: for source characterization, a rate between 1-100 Hz may suffice for coincidence-based measurements \cite{chen2024coincidence}; in contrast, applications relying on non-Gaussian features, such as quantum computing and quantum error correction, require kHz-scale rates for scalability \cite{banic2024quantum}. We also point out that our design principles are not limited to fiber platforms and can be applied to integrated waveguides, where the use of materials with higher nonlinear susceptibilities or structures with smaller effective mode areas can enhance photon triplet generation rates while preserving separability. Although the generation rate presents a limiting factor for practical implementations, we also emphasize that achieving high purity and separability of the triplet state is also critical for applications.

\subsection{\label{sec:resonators_results}Ring resonators}

\begin{figure}[!t]
    \centering
    \includegraphics[width=0.23\textwidth]{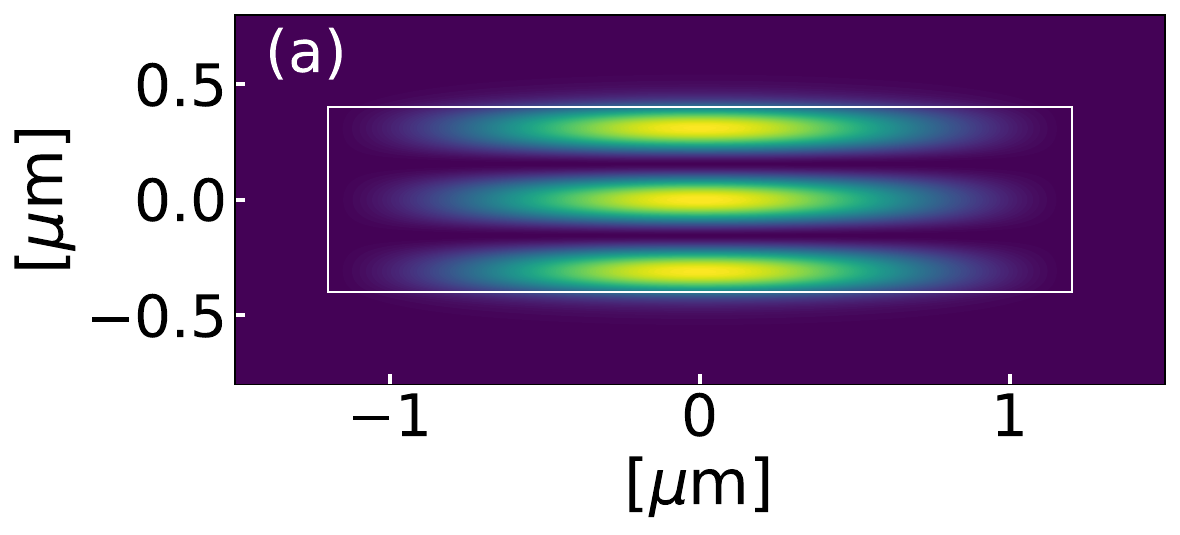}%
    \includegraphics[width=0.23\textwidth]{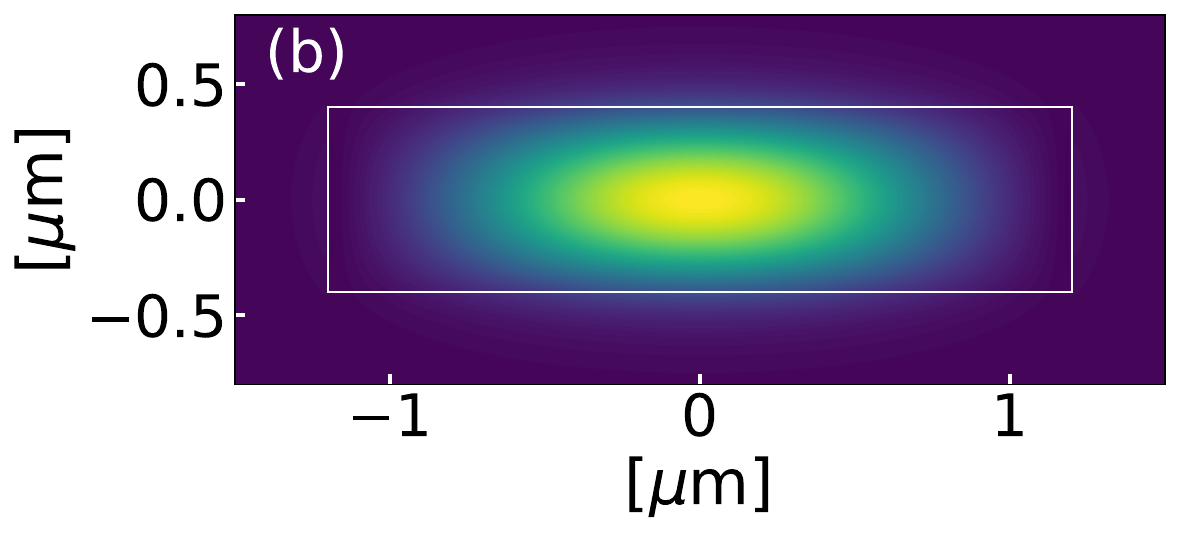}\\
    \vspace{1em}
    \includegraphics[width=0.4\textwidth]{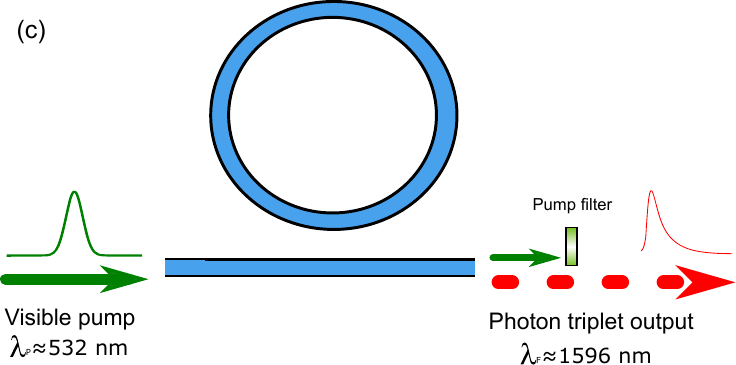}
    \caption{(a) \& (b) Electric field intensities $|\boldsymbol{E}(x,y)|^2$ of the phase matched modes for TOPDC in the sample microring structure. (c) Schematic representation of photon triplet generation in a microring resonator.}
    \label{fig:ring_modes}
\end{figure}

We consider a silicon nitride microring resonator clad in silica, similar to the device discussed in Ref. \cite{banic2022resonant}. We take the ring radius to be 120 $\mu$m, and the cross-sectional dimensions to be 0.8 $\mu$m $\times$ 2.4 $\mu$m. TOPDC in this structure relies on intermodal phase matching between the fundamental and third-order TE modes at the triplet and pump frequencies, respectively (see Fig. \ref{fig:ring_modes}); for the cross-sectional dimensions given above, phase matching is achieved for $\lambda_P \approx 532$ nm.
As in earlier work \cite{banic2022resonant}, we take $Q_F = 10^7$ for the fundamental mode \cite{ElDirani:19}, and $Q_P = 10^5$ for the third-order mode \cite{Surya:18}. The disparity between the two quality factors occurs due to different propagation losses for the pump and fundamental modes.

\begin{figure}[!t]
    \centering
    \includegraphics[width=0.7\linewidth]{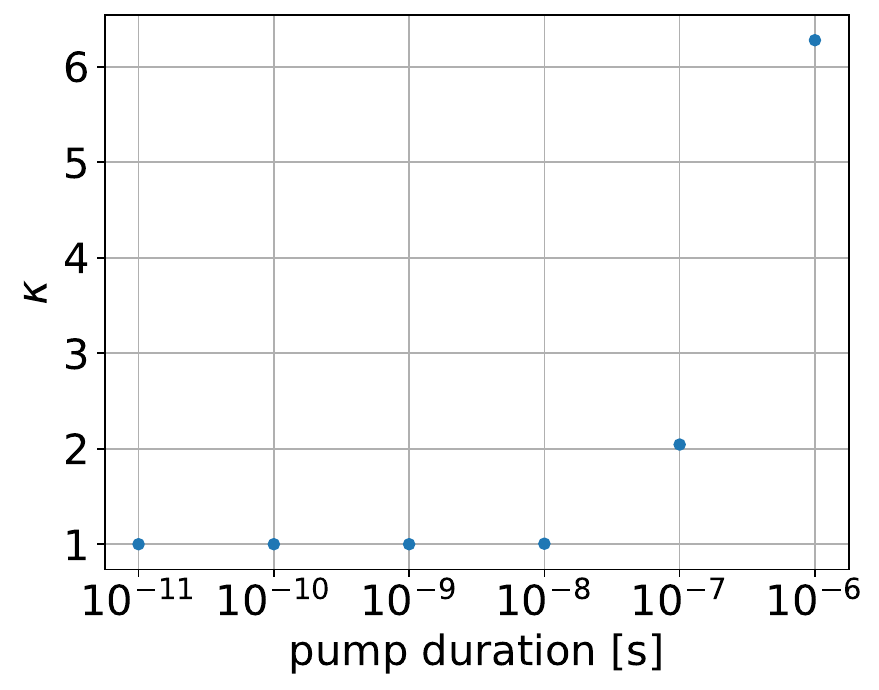}
    \caption{Separability parameter $\kappa$ for a microring resonator with mismatched quality factors ($Q_P = 10^5$, $Q_F = 10^7$), and varying pump pulse durations.}
    \label{fig:kappa_equalQ}
\end{figure}

\begin{figure}[!t]
    \centering
    \includegraphics[width=0.23\textwidth]{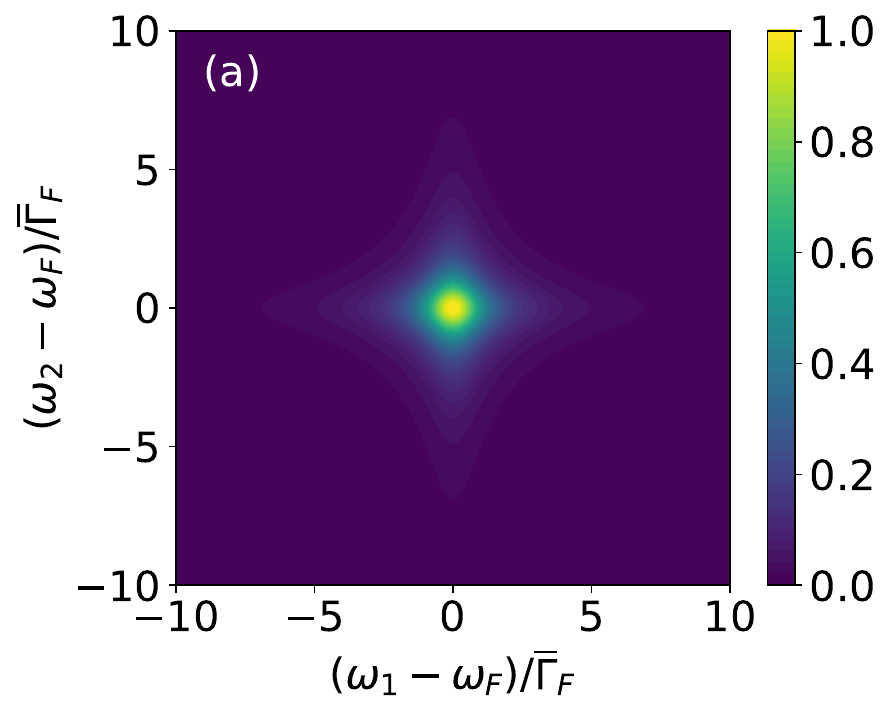}
    \includegraphics[width=0.23\textwidth]{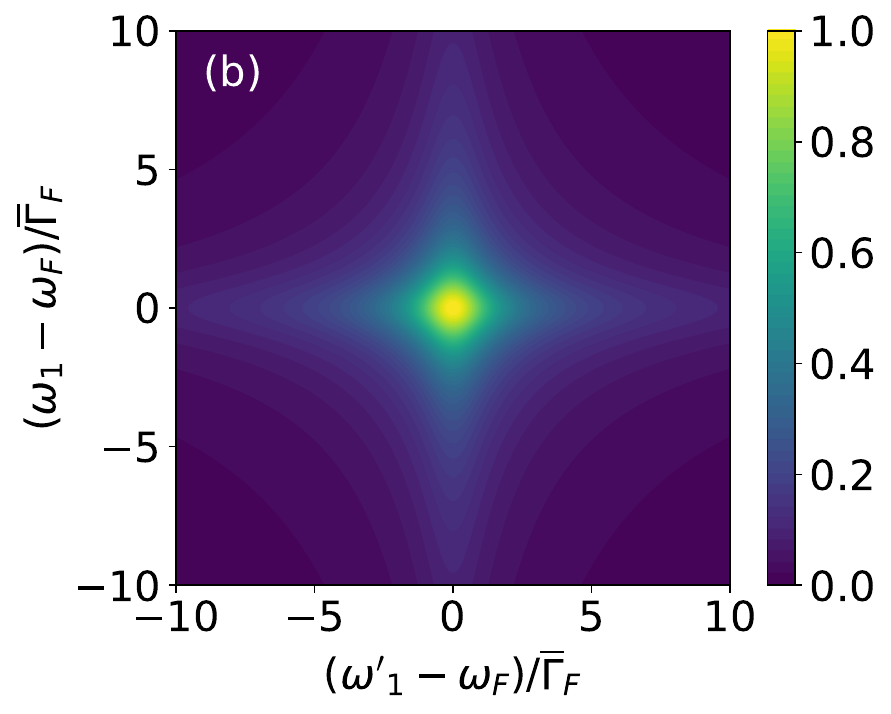}

    \includegraphics[width=0.23\textwidth]{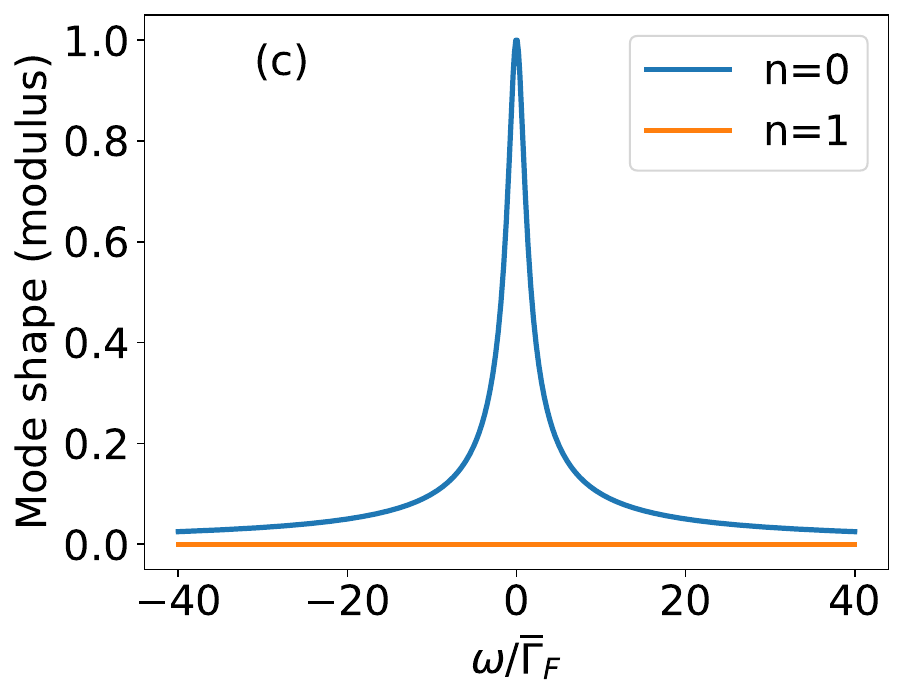}
    \includegraphics[width=0.24\textwidth]{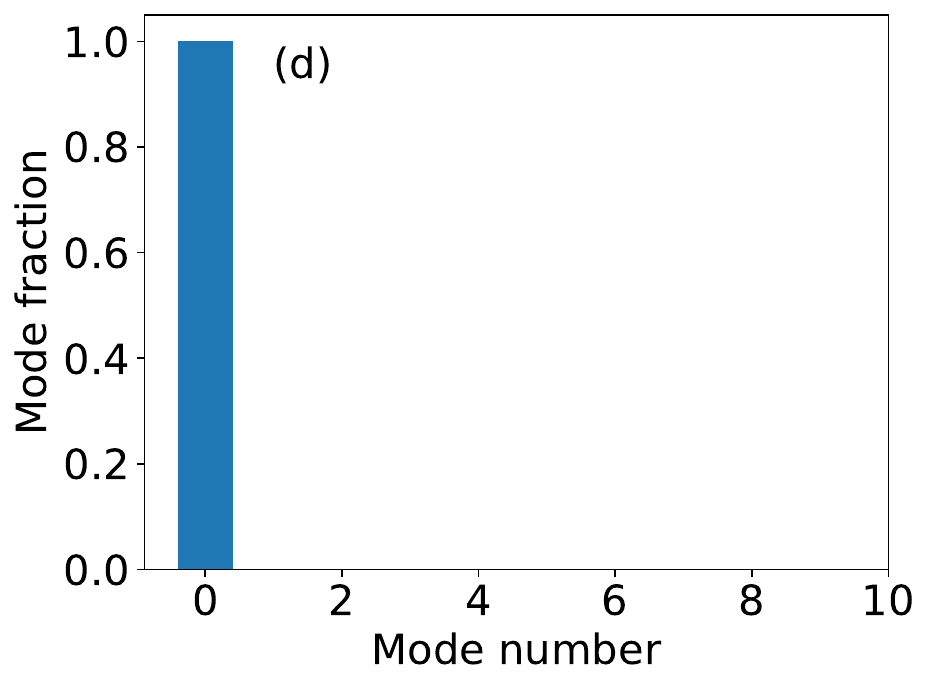}
    \caption{(a) Projection of the JSI $s(\omega_1, \omega_2)$ and (b) single-mode density matrix $\rho(\omega_1, \omega_1')$ for TOPDC in the microring described in Fig. \ref{fig:ring_modes}, with a 1 ns pump duration, and mismatched quality factors for the two transverse modes ($Q_F = 10^7$, $Q_P = 10^5$). In panels (c) and (d), we represent the mode decomposition of the single-mode density matrix.}
    \label{fig:mismatched_Qs}
\end{figure}

In Fig. \ref{fig:kappa_equalQ} we plot $\kappa$ as a function of input pump pulse duration; as expected, we find that $\kappa$ approaches unity as the duration of the pump pulse is reduced and the JSA approaches the separable form in Eq. \eqref{eq:phi_R_sep}. 
In Fig. \ref{fig:mismatched_Qs}(a) we plot the JSI projection $s(\omega_1,\omega_2)$, assuming a $1$ ns pump pulse. In Fig. \ref{fig:mismatched_Qs} (b), we plot the corresponding single-photon density matrix, and 
we represent its mode decomposition in panels (c) and (d). As expected, we find that the triphoton JSA is separable to very good approximation, with the spectral decomposition of $\rho_I$ consisting of a single mode with the Lorentzian form we expect (recall Eq. \eqref{eq:lorentz_mode}). For these parameters we obtain $\kappa - 1 \sim 10^{-6}$. We also see that $s(\omega_1,\omega_2) \approx |\rho(\omega_1,\omega_2)|^2$, as expected for a separable triphoton JSA (recall Eqs. \eqref{eq:reduced_rho} and \eqref{eq:jsa_proj_def}):
If $\Psi(\omega_1,\omega_2,\omega_3) = \psi(\omega_1) \psi(\omega_2) \psi(\omega_3)$, then Eq. \eqref{eq:reduced_rho} becomes
\begin{align}
    \rho(\omega_1,\omega_1') &= \int d\omega_2 d\omega_3 \psi(\omega_1) \psi^*(\omega_1')|\psi(\omega_2)|^2 |\psi(\omega_3)|^2 \\
    &= \psi(\omega_1)\psi^*(\omega_1'),
\end{align}
and Eq. \eqref{eq:jsa_proj_def} takes the form 
\begin{align}
   s(\omega_1,\omega_2) &= \int d\omega_3 |\psi(\omega_1)|^2 |\psi(\omega_2)|^2  |\psi(\omega_3)|^2,\\
   &= |\psi(\omega_1)|^2 |\psi(\omega_2)|^2.
\end{align}

\begin{figure}[!t]
    \centering
    \includegraphics[width=0.23\textwidth]{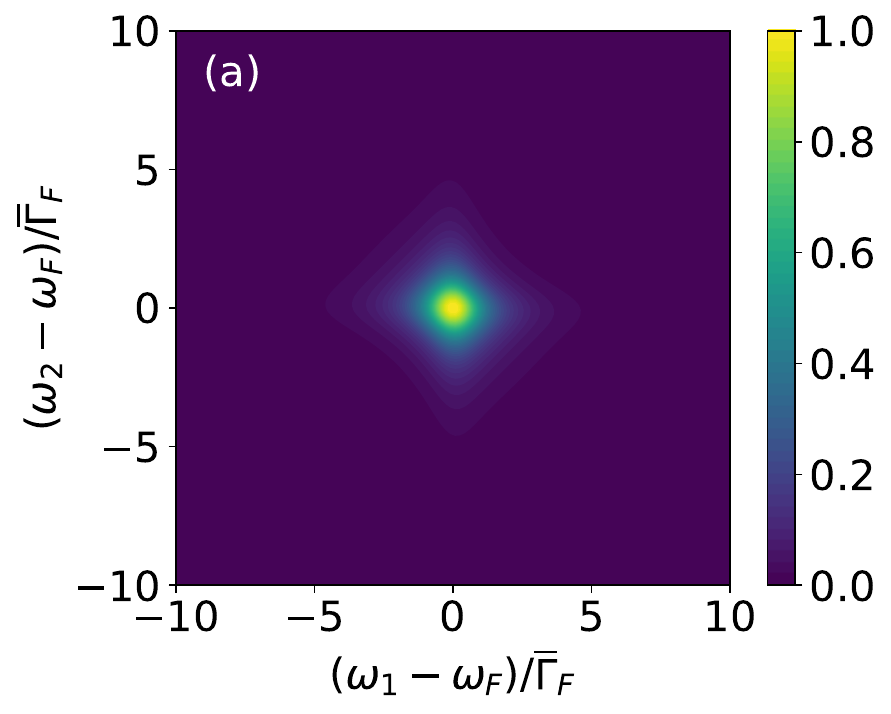}
    \includegraphics[width=0.23\textwidth]{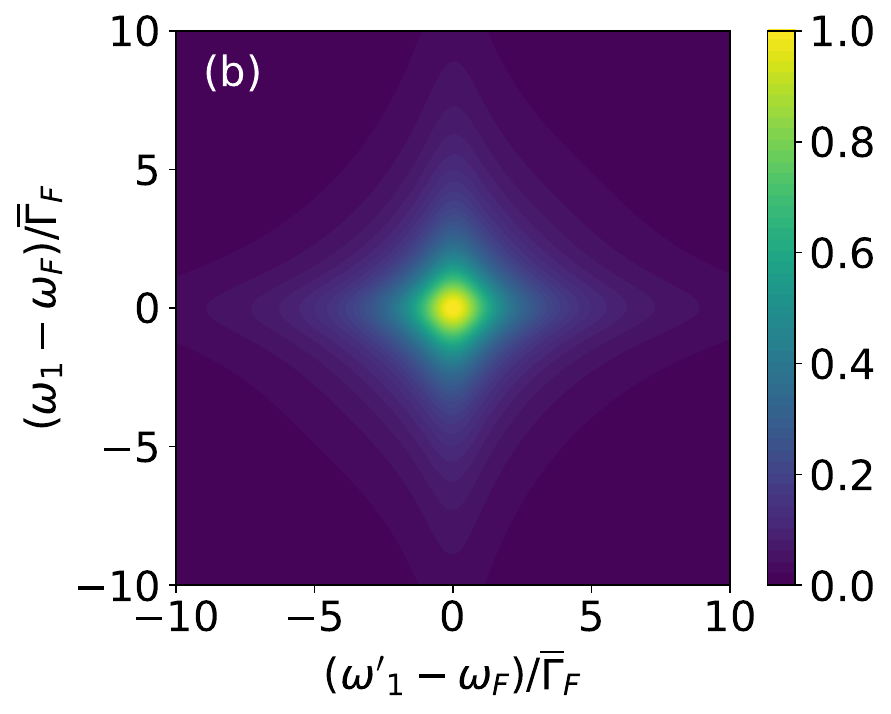}

    \includegraphics[width=0.23\textwidth]{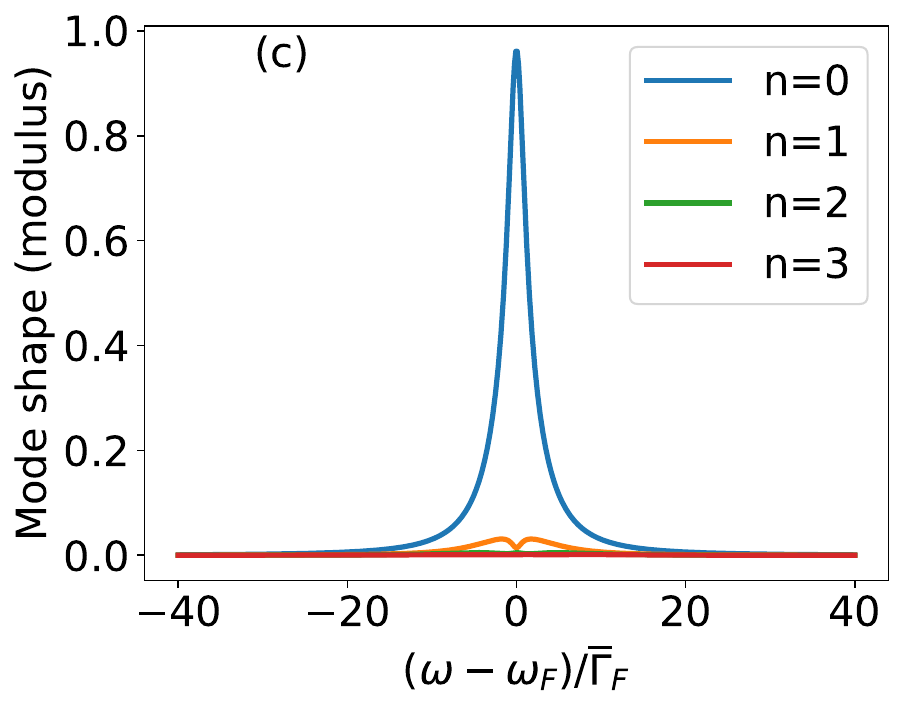}
    \includegraphics[width=0.24\textwidth]{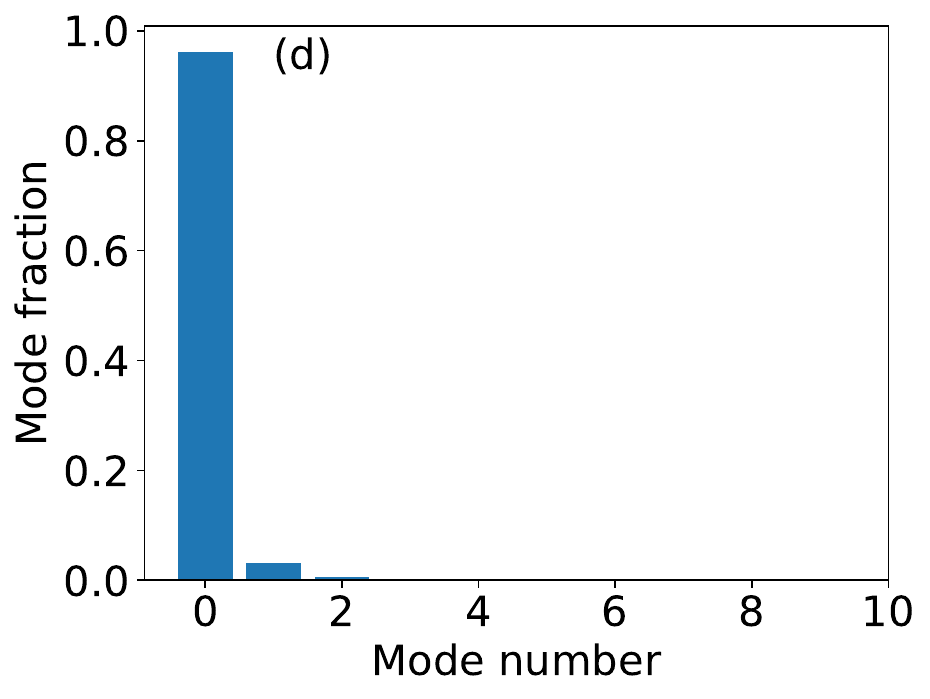}
    \caption{ (a) Projection of the JSI $s(\omega_1, \omega_2)$ and (b) single-mode density matrix $\rho(\omega_1, \omega_1')$ for TOPDC in the microring described in Fig. \ref{fig:ring_modes}, with a 1 ns pump duration, and equal quality factors $Q_F = Q_P = 10^7$. In panels (c) and (d), we represent the mode decomposition of the single-mode density matrix.}
    \label{fig:matchedQ}
\end{figure}

A disparity between the quality factors -- in particular, $Q_F > Q_P$ -- is required to maximize the separability of the JSA \cite{Vernon:17}. Here we have assumed this disparity is achieved ``automatically", due to different propagation losses for the pump and fundamental mode \cite{Surya:18}. If this were not the case, a mismatch in the quality factors could be implemented using interferometric couplers, for example \cite{Vernon:17, Liu:20}. We also point out that even when $Q_F = Q_P$, one can achieve a higher degree of separability compared to the fiber designs discussed above (see Fig. \ref{fig:matchedQ}); the separability saturates around $\kappa = 1.002$ for pulses of a nanosecond or less.

In principle, single-mode triplet generation can be achieved relatively easily using resonant structures, yet this approach is challenging due to the low generation rates expected for typical microring resonators \cite{banic2022resonant}. We calculate the triplet rate as outlined in Appendix \ref{sec:quantum_state_ring}, taking the parameters used in Fig. \ref{fig:mismatched_Qs}, and assuming a nanosecond pump pulse with $\sim10$ pJ pulse energy. The low pulse energy is chosen to ensure the validity of our calculations, which neglect SPM and XPM (see Appendix \ref{sec:quantum_state_ring}). For these parameters, the probability of generating a triplet is $|\varepsilon|^2 \sim 10^{-13}$; assuming a 10 MHz repetition rate, this corresponds to $10^{-6}$ triplets per second, orders of magnitude lower than the rate predicted for the fiber considered above. The triplet rate could be improved by increasing the pump power -- provided the effects of SPM and XPM were taken into account and mitigated \cite{PhysRevApplied.23.054045,PhysRevA.111.063502} -- but because the TOPDC rate scales linearly with pump power, this strategy for triplet generation will likely remain unrealistic.

\section{\label{sec:detection}Detection}
In this section, we describe close-to-optimal strategies to measure the interesting features of a (nearly) spectrally uncorrelated triphoton state.

For direct detection, one would like to detect threefold coincidences after splitting the beam containing the triplets using a three-port interferometer.
We imagine an interferometer in which the first port is fed with light in the state $\ket{\phi}$, and the other two with vacuum. If we label the input creation operator as $a_{F,0}(\omega)$, where $0$ indicates the first input port of the interferometer, then we want to maximize the component with 3 photons in 3 separate ports after making the substitution $a_{F,0}(\omega) \to u_0 a_{F,0}(\omega)+u_1 a_{F,1}(\omega)+u_2 a_{F,2}(\omega)$ in the expression giving the triplet (see Eq. \eqref{eq:psi2}). The component with three photons in three separate ports is simply $6|u_{0} u_1 u_2|^2$, and since the quantities $u_i$ need to be the entries of the column of a unitary matrix (thus $|u_0|^2+|u_1|^2+|u_2|^2 = 1$), the product is maximized when $|u_i|=1/\sqrt{3}$. This implies that the probability of detecting three-fold coincidences, in an idealized scenario with zero losses, is $P_{1,1,1}=6(|\tfrac{1}{\sqrt{3}}|^2)^3 = \tfrac{2}{9}$ of the probability of generating a degenerate triplet. This interferometer can be implemented, for instance, as a balanced tritter~\cite{zukowski1997realizable, campos2000three,spagnolo2013three}.

 Once coupling and detector inefficiencies are considered, the true coincidence detection rate is $R_{\text{real}}=\frac{2}{9} \eta^3 R_{\text{triplet}}$, where the factor 2/9 accounts for the probabilistic splitting of the triplets in the three-fold coincidence scheme described above, $R_{\text{triplet}}$ is the triplet generation rate, and $\eta=\eta_{\text{detector}}\eta_{\text{coupling}}$ is the total quantum efficiency \cite{hammer2018dispersion}. Here, we consider state-of-the-art superconducting nanowire single-photon detectors  (SNSPD) with $\eta_{\text{detector}}=0.95$ \cite{ID281} and coupling efficiency $\eta_{\text{coupling}}\approx 0.5$. For our fiber system with the highest triplet rate, this yields $R_{\text{real}}\approx 8.5$ events/hour, and for the microring system, this yields $R_{real}\approx 8.5\times10^{-5}$ events/hour. 

To address the low triplet generation rate, the coincidence detection scheme can be optimized, for example, using SNSPD with ultra-low dark count rates and high detection efficiencies for triplet coincidence measurements. Reducing the coincidence detection window might also help reduce the number of accidental dark counts. For reference, the accidental triplet coincidence rate due to dark counts is $R_{\text{acc}}\approx R^3_{\text{dark}} \Delta t^2$ \cite{hammer2018dispersion}, where $R_{\text{dark}}$ is the detector's dark count rate and $\Delta t$ is the temporal coincidence window. For the SNSPD systems considered here \cite{ID281}, $R_{\text{dark}}\approx 100~s^{-1}$ and  $\Delta t\approx 100 $ps; this yields around  $3.6\times10^{-11} $ accidental counts/hour.

For homodyne detection with a local oscillator with the spectrum $g(\omega)$, one finds the probability density function (PDF) for the quadrature $x$ at phase $\theta$ to be (cf. 
Appendix~\ref{sec:quadratures})
\begin{align}
p(x|\theta) = \frac{1}{\sqrt{\pi}} e^{- x^2} \left[ 1+\varepsilon \eta \frac{2  }{\sqrt{3}} (-3 x + 2 x^3) \cos 3\theta \right],
\end{align}
where we have neglected terms of order $\varepsilon^2$, and assumed that the mode overlap
\begin{align}\label{eq:eta}
\eta = \int d\omega_1 d\omega_2 d\omega_3 \Psi(\omega_1,\omega_2,\omega_3) g^*(\omega_1) g^*(\omega_2) g^*(\omega_3)
\end{align}
is real, without loss of generality. 
We adopt units in which the shot noise set by the variance of the vacuum equals $\tfrac12$, and we find that  -- to leading order, ignoring $O(\varepsilon^2)$ terms -- of the first four central moments, only the third differs from the values expected from a zero-mean normal distribution; in particular, we have $\braket{x_\theta^3 } = \sqrt{3} \varepsilon \eta \cos 3 \theta$ \cite{chang2020observation}.

To maximize the effects of the triplet component $\ket{\text{III}}$ in the PDF above, and to maximize any deviation from Gaussian statistics in any quadrature measurement, it is necessary to optimize the local oscillator overlap $\eta$. As we show below, near-optimal local oscillators can be obtained from the pseudo-Schmidt functions introduced in Eq.~\eqref{eq:dm}. 

To identify the optimal local oscillator for the detection of triplets, we implement two distinct optimization routines: gradient descent and basin hopping algorithms. Both methods are minimization algorithms; in this case, maximizing $\eta$ is equivalent to minimizing $-\eta$. Gradient descent updates the parameters of $g(\omega)$ in the direction of the gradient of $-\eta$. Convergence is reached when the gradient norm $||\partial_g \eta||$ is sufficiently small, indicating a local optimum \cite{nocedal2006quasi}. In contrast, basing hopping is a global optimization method. At each iteration, a local minimization is performed, followed by a small random perturbation of the solution. This allows the algorithm to escape and refine local minima until a global optimum is found \cite{wales1997global}. In both methods, the final gradient norm $||\partial_g \eta||$ serves as an indicator of convergence. 

For this analysis, we examine two types of triplet sources: An idealized source (as discussed in Section \ref{sec:dispersion_engineering}), and the optimized high-index-contrast optical fibers, previously introduced in Section \ref{sec:optical_fibers}. We define random seeds sampled from a uniform distribution as an initial guess for the function $g(\omega)$. Regardless of the initial seed or the method, the optimization consistently converges to a local oscillator distribution that closely resembles the dominant mode of the single-photon reduced density matrix $\rho(\omega_1,\omega_1')$ (introduced in Section \ref{sec:purity}), for both study cases (see Fig. \ref{fig:jsirho_ideal} (c) and \ref{fig:jsirho} (c)). In Tables \ref{tab:detection1} and \ref{tab:detection2}, we summarize the overlap values $\eta$ and the final gradient norms $||\partial_g \eta||$ obtained for both source types. The results are shown for three local oscillator modes: the dominant mode $f_0(\omega)$ obtained from the spectral decomposition of $\rho(\omega_1,\omega_1')$, the optimal distribution from the gradient descent method $g_{gd}(\omega)$, and the optimal distribution from the basin hopping method $g_{bh}(\omega)$. We found a maximum overlap of $\eta$ $\approx$ 0.92 for the ideal source and $\eta$ $\approx$ 0.87 for the high-index-contrast fibers. These results demonstrate the practical utility of the pseudo-Schmidt modes in designing the local oscillator in the homodyne detection of photon triplets.

\begin{table}[hbtp]
\caption{\label{tab:detection1}%
Local oscillator optimization: Ideal source}
\begin{ruledtabular}
\begin{tabular}{cddd}
g&\multicolumn{1}{c}{\textrm{$\eta$}}&
\multicolumn{1}{c}{\textrm{$||\partial_g \eta||$}}\\
\hline
$f_0(\omega)$&0.92&2.45\mbox{$\times 10^{-2}$}\\
$g_{gd}(\omega)$&0.92& 3.07\mbox{$\times 10^{-7}$}\\
$g_{bh}(\omega)$&0.92  & 7.99\mbox{$\times 10^{-4}$}  \\
\end{tabular}
\end{ruledtabular}
\end{table}

\begin{table}[hbtp]
\caption{\label{tab:detection2}%
Local oscillator optimization: High-index-contrast optical fibers}
\begin{ruledtabular}
\begin{tabular}{cccc}
g&\multicolumn{1}{c}{\textrm{$\eta$}}&
\multicolumn{1}{c}{\textrm{$||\partial_g \eta||$}}\\
\hline
$f_0(\omega)$&0.87&1.40\mbox{$\times 10^{-1}$}\\
$g_{gd}(\omega)$&0.87& 2.27\mbox{$\times 10^{-7}$}  \\
$g_{bh}(\omega)$&0.87  & 1.18\mbox{$\times 10^{-4}$}  \\
\end{tabular}
\end{ruledtabular}
\end{table}

\section{\label{sec:conclusions} Conclusion}

We have analyzed the spectral correlations of photon triplets generated through third-order spontaneous parametric down-conversion (TOPDC) in photonic devices, and we have discussed strategies to quantify and minimize them. To this end, we introduced a measure based on the spectral decomposition of the single-photon reduced density matrix of the triplet state \cite{ma2011measure}. Based on this, we defined a separability parameter $\kappa$, analogous to the Schmidt number in the bipartite case. Minimizing this separability parameter -- with an ideal value of $\kappa = 1$, corresponding to a separable joint spectral amplitude (JSA) -- enables the few-mode operation of the triplet source, which is a fundamental requirement for its application in quantum technologies.   

We discussed two experimental strategies to reduce spectral correlations in TOPDC sources. The first focuses on dispersion engineering in waveguides: We find that when the conditions outlined in Section \ref{sec:detection} are achieved, one obtains an approximately factorable JSA of the triplet state, with the separability parameter reaching a minimum value of $\kappa = 1.25$. The second strategy involves pump engineering in microring resonators. We showed that a factorable JSA -- with $\kappa - 1 \sim 10^{-3}$ --  can be achieved using a pump pulse shorter than the dwelling time of the generated modes, with an even higher degree of separability reached by appropriately engineering the resonance linewidths.

We then applied these strategies to realistic devices. We discussed the optimization of high-index-contrast optical fibers, obtaining a minimum separability parameter of $\kappa = 1.5$, with the few-mode operation maintained even with pump detuning up to 10 nm. The expected triplet generation rate for these fibers is approximately 0.01  triplets per second, taking an average pump power of 50 mW and a 20 cm long fiber. This rate can be increased by raising the pump power or extending the fiber length, provided that SPM and XPM effects are mitigated. While these rates are low and present a detection challenge,  they approach the sensitivity limit of state-of-the-art single-photon detectors. The design principles presented here are not limited to fiber platforms and can be applied to integrated waveguides, where selecting materials with higher nonlinearities or structures with smaller effective mode areas can further enhance triplet generation rates, while preserving spectral separability. We also considered a silicon nitride microring resonator, clad in silica and phase-matched for a 532 nm pump. We found that the separability parameter of the generated triplet state approaches unity as the pump pulse duration is reduced below the resonator dwelling time, indicating a high degree of separability that surpasses that achievable in dispersion-engineered fibers. However, this advantage comes at the cost of extremely low generation rates under typical conditions, making practical implementation challenging.

Finally, we discussed the characterization of nearly uncorrelated triphoton states. We addressed both direct detection, using a three-port interferometer to measure triple coincidences, and homodyne detection. In the latter case, it is essential to properly shape the spectral profile of the local oscillator to maximize deviations from Gaussian statistics in quadrature measurements. We found that the optimal local oscillator corresponds to the dominant mode in the spectral decomposition of the single-photon reduced density matrix. Indeed, achieving single-mode operation in triplet sources will be essential to demonstrate the generation of non-Gaussianity in this way, and we believe that dispersion-engineered non-resonant sources -- like the high-index-contrast fibers discussed in our sample calculations -- are promising candidates to this end.

\section{Data availability statement}

The simulation data and algorithms supporting the findings of this study are available in the following repositories:  \href{https://github.com/polyquantique/separable_triplets}{https://github.com/polyquantique/separable\_triplets} and \href{https://doi.org/10.5281/zenodo.15801470}{https://doi.org/10.5281/zenodo.15801470}.

\begin{acknowledgments}
G.L.O. and N.Q. acknowledge support from the MEI du Québec, EU's Horizon under agreement 101070700 project MIRAQLS, and NSERC, and they thank Rodrigo Becerra-Deana, Karthik Chinni, Martin Houde, and Salvador Poveda-Hospital for insightful discussions. M.B. acknowledges support from the Quantum Research and Development Initiative, led by the National Research Council Canada, under the National Quantum Strategy, and thanks J.E. Sipe for helpful discussions. \end{acknowledgments}

\newpage

\bibliography{fm_triplets_references}
\newpage

\appendix
\section{\label{sec:quantum_state_both} The quantum state of photon triplets generated via TOPDC }

\subsection{\label{sec:quantum_state_wg} Waveguides }
The interaction Hamiltonian that describes the TOPDC interaction in waveguides is \cite{banic2024quantum}:
\begin{align} \label{eq:Hinter}
    \hat{H}^{(I)}_{FFFP}=-\int d\omega_1 d\omega_2 d\omega_3 d\omega_P  M_{FFFP}(\omega_1, \omega_2,\omega_3,\omega_P) \nonumber \\  a^\dagger_F(\omega_1)a^\dagger_F(\omega_2) a^\dagger_F(\omega_3) e^{-i \Delta \omega t}+\text{H.c.},
\end{align}
where $\Delta \omega=\omega_P-\omega_1-\omega_2-\omega_3$, and 
\begin{align}\label{eq:Mfffp}
    M_{FFFP}(\omega_1, \omega_2,\omega_3,\omega_P)=\frac{\hbar^2\overline{\omega}_F}{12 \pi^2} \gamma_{FFF}~\ell~ \alpha \left(\omega_P\right)\varphi\left(\frac{\ell}{2} \Delta k \right).
\end{align}
$\hbar$ is the reduced Planck constant, $\overline{\omega}_F$ is the central frequency of the photon triplets, $\ell$ is the waveguide length, and $ \gamma_{FFF}$ is the nonlinear parameter, which quantifies the efficiency in the generation of the photon triplets \cite{banic2022resonant}; $\alpha \left(\omega_P\right)$ is the pump envelope function, normalized such that $\int d\omega |\alpha(\omega)|^2=N_P$, where $N_P$ is the average number of pump photons, and $\varphi\left(\Delta k ~\ell/2\right)$ is the phase-matching function.  
For waveguides homogeneous in the longitudinal direction, $\varphi\left( \Delta k~ \ell / 2 \right)=\sinc\left( \Delta k ~\ell / 2 \right)$. In \eqref{eq:Mfffp}, we assumed that the pump was prepared in a strong coherent state, such that it remains undepleted in the propagation through the waveguide. We have also ignored other phase-matched nonlinear optical effects, like self and cross-phase modulation, but they can be easily included.  

In perturbation theory, the quantum state resulting from the TOPDC generation in waveguides is calculated via
\begin{equation}\label{eq:dynamics}
    \ket{\phi}= \mathcal{T} \exp \left[\frac{-i}{\hbar}\int_{-\infty}^{\infty} dt \hat{H}^{(I)}_{FFFP} (t)\right] \ket{\text{vac}},
\end{equation} 
in the spontaneous regime, in which the triplet modes are initially in the vacuum state. $\mathcal{T}$ is the time-ordering operator. 
Considering a low-gain interaction, the ket can be approximated to first order as:
\begin{equation}\label{eq:psi_4}
    \ket{\phi} \approx \ket{\text{vac}}+\varepsilon \ket{\text{III}},
\end{equation}
where
\begin{eqnarray} \label{eq:psi3-4}
    \ket{\text{III}}=\frac{1}{\sqrt{6}} \int d\omega_1 d\omega_2 d\omega_3 ~& &\Psi (\omega_1,\omega_2,\omega_3)\nonumber \\ & &a^\dagger_F(\omega_1)a^\dagger_F(\omega_2) a^\dagger_F(\omega_3)\ket{\text{vac}}. \nonumber \\
\end{eqnarray}
$\Psi  (\omega_1,\omega_2,\omega_3)$ is the triphoton JSA, and $|\varepsilon|^2$ is the probability of generating a photon triplet per pump pulse.
\begin{align} \label{eq:beta}
    |\varepsilon|^2=&\frac{1}{6 \pi^2} |\gamma_{FFF}|^2 \ell^2 \hbar^2 \omega_F^2  \nonumber \\ &\int d\omega_1 d\omega_2 d\omega_3 ~ \alpha^2(\omega_1,\omega_2,\omega_3)~\sinc^2\left(\Delta k\frac{\ell}{2}\right) 
\end{align}
Finally,
\begin{align}\label{eq:jsa_a}
    \Psi  (\omega_1,\omega_2,\omega_3)=\frac{i}{\varepsilon \sqrt{6}} & \frac{\hbar \omega_F}{\pi} \gamma_{FFF}~ \ell~  \nonumber \\ \alpha(&\omega_1+\omega_2+\omega_3)~\sinc\left(\Delta k\frac{\ell}{2}\right).
\end{align}

\subsubsection{\label{sec:wg_SPM_XPM} Self- and cross-phase modulation}

While the above analysis assumed SPM and XPM to be negligible, in practice, ultrashort pulses can reach powers high enough for these nonlinearities to influence propagation. Both effects arise from the intensity dependence of the refractive index. In our context, these processes can broaden the pump spectra and alter the TOPDC phase-matching condition through an additional nonlinear phase shift \cite{banic2022resonant}. To evaluate the impact of these effects in our designs, we adopt the nonlinear length introduced by \citeauthor{agrawal2019nonlinear} \cite{agrawal2019nonlinear} as a figure of merit. The nonlinear length estimates the characteristic distance over which nonlinear effects become significant. For SPM and XPM, it is defined as:
\begin{equation}
\begin{aligned}
L_{\text{SPM}} &= \frac{1}{\gamma_{\text{SPM}}\,P_{\text{peak}}}
\quad\quad
&
L_{\text{XPM}} &= \frac{1}{\gamma_{\text{XPM}}\,P_{\text{peak}}}
\end{aligned}
\end{equation}

Here, $\gamma_{\text{SPM}}$ and $\gamma_{\text{XPM}}$ are the respective nonlinear parameters as defined in \cite{banic2022resonant}, and $P_{\text{peak}}$ is the pump peak power. For the fibers presented in Section \ref{sec:optical_fibers}, $|\gamma_{\text{SPM}}|\approx 0.032$ (Wm)$^{-1}$ and $|\gamma_{\text{XPM}}|\approx 0.010$ (Wm)$^{-1}$. We choose $P_{\text{peak}}$ such that the fiber length $\ell$ remains greater or equal than the corresponding nonlinear lengths, thus keeping SPM and XPM within acceptable limits. Considering $\ell=20$ cm, the maximum peak power allowed by SPM is 156.25 W. In our proposal, we take a pump average power of 50 mW and  pump pulse duration of 31 fs, this corresponds to repetition rates on the order of 9.7 GHz, for which $P_{\text{peak}}$ remains safely below the SPM threshold. Such GHz femtosecond pulsed lasers have been demonstrated in the literature \cite{sekhar202320,chapman2013femtosecond}.

\subsection{\label{sec:quantum_state_ring} Microring resonators}

The derivation of the state generated by TOPDC in a resonator is analogous to the one described in Section \ref{sec:quantum_state_wg}, but putting 
\begin{align}
    M_{FFFP}(\omega_1, \omega_2, \omega_3, \omega_P) &= \frac{\hbar^2 \overline{\omega}_F}{12 \pi^2} \gamma_{FFF} \mathcal{L} F^*_{F+}(\omega_1) F^*_{F+}(\omega_2) \nonumber \\ &\times F^*_{F+}(\omega_3) F_{P-}(\omega_P) \alpha(\omega_P), \label{eq:M_ring_def}
\end{align}
where $\mathcal{L}$ denotes the circumference of the microring, and $F_{J\pm}(\omega)$ denotes the field enhancement factor given in Eq. \eqref{eq:F_def}; the other parameters are defined in Section \ref{sec:quantum_state_wg}. Putting Eq. \eqref{eq:M_ring_def} into \eqref{eq:Hinter} and proceeding as described in Section \ref{sec:quantum_state_wg}, one obtains the JSA given in Eq. \eqref{eq:JSA_ring}, and 
\begin{align}
    |\varepsilon|^2 &=\frac{\hbar^2\overline{\omega}_F^2}{6\pi^2}|\gamma_{FFF}|^2 \mathcal{L}^2 \int d\omega_1 d\omega_2 d\omega_3 |F_{F+}(\omega_1)|^2 |F_{F+}(\omega_2)|^2 \nonumber \\ &\times |F_{F+}(\omega_3)|^2 |F_{P-}(\omega_1+ \omega_2 + \omega_3)|^2 |\alpha(\omega_1 + \omega_2 + \omega_3)|^2.  
\end{align}

\subsubsection{Self- and cross-phase modulation}

The effect of SPM in a resonant source can be understood as causing a shift in the resonance frequency \cite{banic2022resonant}. This can be understood by considering the evolution of a ring mode operator $b_J$ under $ H_\text{ring} + H_{SPM} + H_{XPM}$ \cite{beyondphotonpairs}. For each mode $J$, the term in the equation of motion due $H_{\text{ring}}$ is 
\begin{align}
    -\frac{i}{\hbar}[b_J(t),H_{\text{ring}}(t)]
    &= -i \omega_J b_{J}(t). \label{eq:ring_free}
\end{align}
We focus first on SPM on the pump field; we have
\begin{align}
    H_{SPM} = -\frac{1}{2} \hbar^2 \omega_P \frac{1}{\mathcal{L}} \gamma_{SPM} v_P^2 b^{\dagger}_P b^{\dagger}_P b_P b_P,
    \label{eq:HSPM_ring}
\end{align}
with the nonlinear parameter defined in the usual way \cite{beyondphotonpairs, banic2022resonant}. The ring operator associated with the pump resonance then evolves under $H_{SPM}$ according to

\begin{align}
    -\frac{i}{\hbar}[b_P(t),H_{SPM}(t)]
    &= i\hbar \omega_P \frac{1}{\mathcal{L}} \gamma_{SPM} v_P^2 b^{\dagger}_{P}(t) b_{P}(t) b_{P}(t),\\
    &= i\hbar \omega_P \frac{1}{\mathcal{L}} \gamma_{SPM} v_P^2 |\beta_P(t)|^2 b_{P}(t). \label{eq:ring_SPM1},\\
    &= i \gamma_{SPM} v_P P'_P(t) b_{P}(t). \label{eq:ring_SPM2}
\end{align}
In (\ref{eq:ring_SPM1}) we have taken $<b^{\dagger}_P(t) b_P(t)> = |\beta_P(t)|^2$; in (\ref{eq:ring_SPM2}) we introduce $P'_P(t) = \hbar \omega_P \frac{1}{\mathcal{L}} v_P |\beta_P(t)|^2$, the power of the pump field in the ring. Therefore the evolution of the fields in the ring including SPM can be written in a form analogous to Eq. \eqref{eq:ring_free}, with a time-dependent resonance frequency: 
\begin{align}
    -\frac{i}{\hbar}[b_J(t),H_{\text{ring}}(t) + H_{SPM}(t)]
    &= -i \tilde{\omega}_J(t) b_{J}(t), \label{eq:ring_timedep}\\
    \tilde{\omega}_J(t) &= \omega_J - v_P \gamma_{SPM} P'(t).
\end{align}

For simplicity, we focus on a parameter regime in which SPM can be neglected. Physically, this corresponds to a regime in which the resonance frequency shift due to SPM is much smaller than the resonance linewidth, i.e. 
\begin{align}
    v_P |\gamma_{SPM}| P'(t) \ll \overline{\Gamma}_P,
\end{align}
where $\overline{\Gamma}_P$ is the half width at half maximum of the resonance. This regime can be achieved by restricting the pump power to satisfy
\begin{align}
    P'(t) &\ll \frac{\overline{\Gamma}_J}{v_P |\gamma_{SPM}|}.
\end{align}
For the microring discussed in the main text, we have $|\gamma_{SPM}| \approx 4 $ (Wm)$^{-1}$ \cite{banic2022resonant}, $Q = 10^5$ which corresponds to $\overline{\Gamma}_P = 16.4$ GHz, and $v_P = 1.3 \times 10^8$ m/s. For these parameters
\begin{align}
    \frac{\overline{\Gamma}_P}{v_P |\gamma_{SPM}|} \approx 30 W,
\end{align}
thus SPM should be negligible if the peak power in the ring is limited to a few watts. The maximum field enhancement factor is 
\begin{align}
    |F(\omega_J)|^2 = \frac{2v_J \eta}{\mathcal{L} \overline{\Gamma}_J}, 
\end{align}
where $\eta$ is the escape efficiency. For critical coupling ($\eta = 0.5$) we have $|F(\omega_P)|^2 = 10$, therefore a peak power of $\sim1$ W in the ring corresponds to an input peak power on the order of $\sim100$ mW. 

A similar analysis can be applied to XPM on the triplet field due to the pump. Here 
\begin{align}
    H_{XPM} = - 2 \hbar^2 {\omega_{F}} \frac{1}{\mathcal{L}} \gamma_{XPM} v_P v_F b^{\dagger}_P b^{\dagger}_F b_P b_F,
\end{align}
and 
\begin{align}
    -\frac{i}{\hbar}[b_F(t),H_{XPM}(t)]
    &= 2i \gamma_{XPM} v_G P'_P(t) b_{F}(t). \label{eq:ring_XPM},
\end{align}
so for the effect of XPM to be negligible, we require 
\begin{align}
     2 v_G|\gamma_{XPM}|  P'_P(t) \ll \overline{\Gamma}_F,\\
     P'_P(t) \ll \frac{\overline{\Gamma}_F}{2 v_G|\gamma_{XPM}|}.
\end{align}
For the fundamental resonance we have $Q = 10^7$, corresponding to $\overline{\Gamma}_P = 54.8$ MHz, $v_F = 1.4 \times 10^8$ m/s, and $|\gamma_{SPM}| \approx 0.8 $ (Wm)$^{-1}$ \cite{banic2022resonant}. We have
\begin{align}
    \frac{\overline{\Gamma}_F}{2 v_F |\gamma_{XPM}|} \approx 200 mW.
\end{align}

We thus take a nanosecond Gaussian pulse of 10 pJ, for which the peak power is $9$ mW, with $\sim 90$ mW circulating in the ring.

\section{\label{sec:quadratic} Quadratic form of the phase mismatch }

Eq. \eqref{eq:delk} defines the level surfaces of the phase matching function (PMF) for constant values of $(\ell/2)\Delta k=C_0$ within the domain of the PMF. If the dispersion relations of the pump and triplet modes can be approximated by a second-order Taylor series expansion around their central frequencies, then the eq. \eqref{eq:delk} describes a set of quadric surfaces. For simplicity, we will assume zeroth-order phase matching $\left(\overline{k}_P-3\overline{k}_F=0 \right)$ and approximate group-velocity matching $(v_P\approx v_F)$. Then, we can write
\begin{equation}
    \frac{\ell}{2}\Delta k=\bm{w}^T \bm{A} \bm{w}=C_0,
\end{equation}
where $\bm{w}=\left(\delta \omega_1,\delta \omega_2,\delta \omega_3\right) $, and $\bm{A}$ is a real symmetric matrix. In our case
\begin{align}  
\bm{A}=\frac{\ell}{4}
\begin{pmatrix}
    \beta_P-\beta_F & \beta_P & \beta_P \\
     \beta_P & \beta_P-\beta_F &\beta_P \\
     \beta_P & \beta_P & \beta_P-\beta_F 
\end{pmatrix}
\end{align}

The eigenvectors of $\bm{A}$ define the directions of the principal axes of the PMF. The eigenvalues $\lambda_i$ of $\bm{A}$ are 
\begin{align}
    \lambda_1 &=\frac{\ell}{4}\left(3\beta_P-\beta_F\right)\\
    \lambda_2 &=\lambda_3=-\frac{\ell}{4}\beta_F,
\end{align}
with associated normalized eigenvectors
\begin{align}
    \bm{v}_1 &=(1,1,1)^T/\sqrt{3}\\
    \bm{v}_2 &=(0,-1,1)^T/\sqrt{2}\\
    \bm{v}_3 &= (2,-1,-1)^T/\sqrt{6}
\end{align}
If the eigenvalues of $\bm{A}$ and $C_0$ are all positive, or all negative, then the PMF isosurfaces correspond to concentric ellipsoids, whose amplitude will depend on the value of $C_0$ in the equation  $\varphi(C_0)= \sinc(C_0)$. For example, the PMF is maximum at the point $\omega_1=\omega_2=\omega_3=\overline{\omega}_F=\overline{\omega}_P/3$ in which $C_0=0$. 

If $|\beta_F| \gg |\beta_P|$, then the isocurves form spherical layers, resulting in a symmetric spectral distribution. In this case, the bandwidth of the PMF is $\sigma_{{\scriptscriptstyle PM}}=\sqrt{4\pi/\ell |\beta_F|}$, defined by the sinc function's central peak at $C_0=\pi$.

\section{\label{sec:quadratures} Quadrature distribution for photon triplets }

Let us introduce a broadband generalized quadrature operator
\begin{equation}
    X_\theta = \frac{A^\dagger e^{i\theta}+A e^{-i\theta}}{\sqrt{2}},
\end{equation}
defined in terms of a broadband bosonic mode  
\begin{equation}
    A^\dagger=\int d\omega_i g(\omega_i) a_F^\dagger(\omega_i),
\end{equation}
and its corresponding $\delta-$normalized eigenstate \cite{barnett1997methods}
\begin{equation}\label{eq:quadrature_state}
    \ket{x_\theta}=\frac{1}{\pi^{1/4}} \exp\left(-\frac{1}{2}x^2+\sqrt{2} e^{i\theta}x A^\dagger-\frac{1}{2} e^{2i\theta}A^{\dagger 2}\right)\ket{\text{vac}}.
\end{equation}
 Using \eqref{eq:psi2} and \eqref{eq:quadrature_state}, we obtain, the quadrature representation of the triplet state $\ket{\phi}$ as
\begin{equation}
    \braket{\phi|x_\theta}=\frac{1}{\pi^{1/4}} e^{-\frac{1}{2}x^2} \left[1+\frac{\varepsilon\eta e^{3i\theta} }{\sqrt{3}}\left(2 x^3-3 x\right)\right], 
\end{equation}
where $\eta$ is the mode overlap
\begin{align}
\eta = \int d\omega_1 d\omega_2 d\omega_3 \Psi(\omega_1,\omega_2,\omega_3) g^*(\omega_1) g^*(\omega_2) g^*(\omega_3).
\end{align}
assumed real without loss of generality.

Finally, the quadrature distribution of photon triplets $p(x|\theta)=|\braket{\phi|x_\theta}|^2$ is equal to
\begin{align}
p(x|\theta) = \frac{1}{\sqrt{\pi}} e^{- x^2} \left[ 1+\varepsilon \eta \frac{2  }{\sqrt{3}} (-3 x + 2 x^3) \cos 3\theta \right],
\end{align}
where we have neglected terms of order $\varepsilon^2$ and assumed the probability amplitude $\varepsilon$ to be real without loss of generality. 
From the distribution above, we can obtain (to leading order in $\varepsilon$) the central moments
\begin{align}
\braket{x_\theta} &= 0, \\
\braket{x_\theta^2} &= \tfrac{1}{2}, \\
\braket{x_\theta^3}&= \varepsilon \eta \sqrt{3} \cos(3 \theta ), \\
\braket{x_\theta^4}&=\tfrac{3}{4}.
\end{align}
Note that the first, second, and fourth moments are identical to those of a zero-mean normal distribution with variance $\frac{1}{2}$. The third-order moment, however, gives information about the non-Gaussianity of the state and is sensitive to the phase of the local oscillator $\theta$.

\end{document}